\documentclass{aa}  
\usepackage{graphicx}
\usepackage{txfonts}
\def\ltapprox{\raise 2pt \hbox {$<$} \kern-1.1em \lower 5pt \hbox {$\approx$}}

\def\ltsim{\; \raise0.3ex\hbox{$<$\kern-0.75em \raise-1.1ex\hbox{$\sim$}}\; }
\def\gtsim{\; \raise0.3ex\hbox{$>$\kern-0.75em \raise-1.1ex\hbox{$\sim$}}\; }

\def\ie{{\it i.e.,~}}

\def\eg{{\it e.g.,~}}
\def\egg{{\it e.g.~}}
\begin{document}
   \title{Unveiling radio halos in galaxy clusters in the LOFAR era}

%   \subtitle{I. Expectations for the 120 MHz MS$^3$ survey}

   \author{R. Cassano\inst{1}\fnmsep\thanks{\email{rcassano@ira.inaf.it}},
   G. Brunetti\inst{1}, H. J. A. R\"ottgering\inst{2}, M. Br\"uggen\inst{3}} 
\authorrunning{R. Cassano et al.}

   \offprints{R.Cassano}

   \institute{INAF - Istituto di Radioastronomia, via P. Gobetti 101,I-40129 Bologna, Italy\\
   \and Leiden Observatory, Leiden University, Oort Gebouw, P.O. Box 9513, 2300 RA Leiden, The Netherlands \\
         \and Jacobs University Bremen, P. O. Box 750 651, 28725, Bremen, Germany \\}

%   \date{Received...; accepted...}

\abstract
{} 
{Giant radio halos are mega-parsec scale synchrotron sources 
detected in a fraction of massive and merging galaxy clusters.
Radio halos provide one of the most important pieces of evidence 
for non-thermal components in large scale structure.
Statistics of their properties can be used to discriminate 
among various models for their origin. 
Therefore, theoretical predictions of the occurrence of radio halos are important as several new radio telescopes are about to begin to survey the sky at low frequencies with unprecedented sensitivity.}
{In this paper we carry out Monte Carlo simulations to model the
formation and evolution of radio halos in a cosmological framework.
We extend previous works on the statistical properties of radio halos 
in the context of the turbulent re-acceleration model.}
{First we compute the fraction of galaxy clusters that show radio halos and derive 
the luminosity function of radio halos.
Then, we derive differential and integrated number count distributions 
of radio halos at low radio frequencies with the main goal to explore
the potential of the upcoming LOFAR surveys.
By restricting to the case of clusters at redshifts $<$ 0.6, 
we find that the planned LOFAR all sky survey at 120 MHz is expected to 
detect about 350 giant radio halos. About half of these halos have spectral indices larger than 1.9 and substantially brighten at lower frequencies.  
If detected they will allow for a confirmation that turbulence accelerates the emitting particles. We expect that also commissioning surveys, such as MS$^3$, have the potential to detect about 60 radio halos in clusters of the ROSAT Brightest Cluster Sample and its extension (eBCS). These surveys will allow us to constrain how the rate of formation of radio halos in these clusters depends on cluster mass.}
{}


\maketitle

\section{Introduction}

Radio halos are diffuse Mpc--scale radio sources
observed at the center of $\sim 30\%$ of massive galaxy clusters 
(\egg Feretti 2005; Ferrari et al. 2008, for reviews).
These sources emit synchrotron radiation due to GeV electrons diffusing through
$\mu$G magnetic fields and provide the most important evidence of non-thermal components in the intra-cluster-medium (ICM). 

Clusters hosting radio halos always show very recent or 
ongoing merger events (\eg Buote 2001; 
Schuecker et al 2001; Govoni et al. 2004; Venturi et al. 2008). This 
suggests a connection between the gravitational process of cluster formation 
and the origin of these halos. Cluster mergers are expected 
to be the most important sources of non-thermal components in 
galaxy clusters. A fraction of the energy dissipated during these mergers 
could be channelled into amplification of the magnetic fields 
(\eg Dolag et al. 2002; Br\"uggen et al. 2005; Subramanian et al. 2006; 
Ryu et al. 2008) and into the acceleration of high energy particles 
via shocks and turbulence (\eg En\ss lin et al. 1998; Sarazin 1999; 
Blasi 2001; Brunetti et al. 2001, 2004; Petrosian 2001; Miniati et al. 2001; 
Fujita et al. 2003; Ryu et al. 2003; Hoeft \& Br\"uggen 2007; 
Brunetti \& Lazarian 2007; Pfrommer et al. 2008, Brunetti et al. 2009).

A promising scenario proposed to explain the origin of the synchrotron 
emitting electrons in radio halos assumes that electrons are 
re-accelerated due to the interaction with MHD turbulence injected 
in the ICM in connection with cluster mergers 
({\it turbulent re-acceleration} model,
\eg Brunetti et al. 2001; Petrosian 2001).
An alternative possibility is that the emitting electrons are continuously injected by {\it pp} collisions in the ICM ({\it secondary} models; \eg Dennison 1980; Blasi \& Colafrancesco 1999).
 
In the picture of the {\it turbulent re-acceleration} scenario, the formation 
and evolution of radio halos are tightly connected with the dynamics and evolution
of the hosting clusters. 
Indeed, the occurrence of radio halos at any redshift depends on 
the rate of cluster-cluster mergers and on the fraction of 
the merger energy channelled into MHD turbulence and re-acceleration 
of high energy particles. 
In the last few years this has been modeled through Monte Carlo 
procedures (Cassano \& Brunetti 2005; Cassano et al. 2006a) that 
provide predictions that can be verified using future instruments.
In this scenario radio halos have a relatively short lifetime
($\approx$ 1 Gyr), and the fraction of galaxy clusters where 
radio halos are generated is expected to increase with cluster mass (or X-ray
luminosity), since the energy of the turbulence generated
during cluster mergers is expected to scale
with the cluster thermal energy (which roughly scales as $\sim M^{5/3}$;
\eg Cassano \& Brunetti 2005).
It has been shown that the predicted 
occurrence of radio halos as a function of the cluster mass (or X-ray
luminosity) is in line with results obtained from a large observational 
project, the ``GMRT radio halo survey'' (Venturi et al.~2007, 2008), 
and with its 
combination with studies of nearby halos based on the NVSS survey
(\eg Cassano et al.~2008).

\begin{figure}
\centerline{
\includegraphics[width=7.3cm,height=6.5cm]{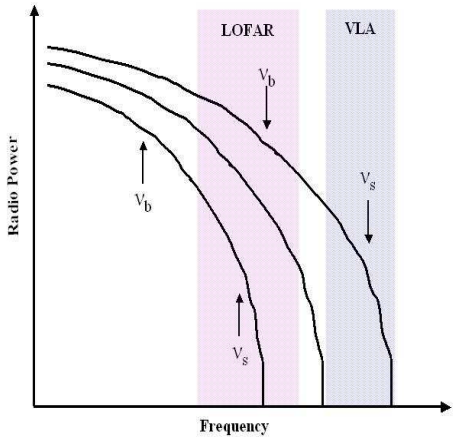}
}
\caption{A schematic representation of the theoretical synchrotron spectra of radio halos with different values of $\nu_s$ (and $\nu_b$). The colored regions indicate the frequency range of VLA and LOFAR observations.} 
\label{fig:ussrh}
\end{figure}

\noindent
The steep spectrum of radio halos makes these sources 
ideal targets for observations at low radio frequencies suggesting
that present radio telescopes can only detect the tip of the iceberg
of their population (En\ss lin \& R\"ottgering 2002; Cassano et al.~2006; Hoeft et al. 2008).
The recent discovery of the giant and ultra-steep spectrum radio halo
in Abell 521 at low radio frequencies (Brunetti et al. 2008) allows a first 
confirmation of this conjecture and provides a glimpse of what future
low frequency radio telescopes, such as the Low Frequency Array (LOFAR)\footnote{http://www.lofar.org} and the Long Wavelength Array (LWA, \eg Ellingson et al. 2009), might find in the upcoming years.

\noindent
LOFAR promises an impressive gain of two orders of magnitude 
in sensitivity and angular resolution over present instruments in 
the frequency range 15--240 MHz, 
and as such will open up a new observational window to the 
Universe. In particular, LOFAR is expected to contribute significantly to the 
understanding of the origin and evolution of the relativistic matter 
and magnetic fields in galaxy clusters.

The main focus of the present paper is to provide a theoretical framework 
for the interpretation of future LOFAR data by quantifying expectations 
for the properties and occurrence of giant radio halos in the context of the 
{\it turbulent re-acceleration} scenario.
In particular, in Sect. 2 we summarize the main ingredients used in
the model calculations and provide an extension of the results of
previous papers on the occurrence of radio halos in
clusters (Sect.~2.1) and on the expected radio halo luminosity 
functions (Sect. 2.2). 
In Sect.~3 we derive the expected number counts of radio halos 
at 120 MHz and explore the potential of LOFAR surveys. 
Our conclusions are given in Sect.4.

\noindent

A $\Lambda$CDM ($H_{o}=70$ $\mathrm{ Km\, s^{-1} Mpc^{-1}}$, $\Omega_{m}=0.3$, 
$\Omega_{\Lambda}=0.7$) cosmology is adopted throughout the paper.

\section{Statistical modelling of giant radio halos in galaxy clusters}

Turbulence generated during cluster mergers may accelerate relativistic
particles and produce diffuse synchrotron emission from Mpc regions
in galaxy clusters (\eg Brunetti et al. 2008).
Diffuse radio emission in the form of 
giant radio halos should be generated in connection with massive
mergers and fade away as soon as turbulence is dissipated and the emitting electrons cool due to radiative losses.
It is likely that the generation of turbulence and the acceleration of 
particles persist for a few crossing times of the cluster-core
regions. This then yields a lifetime of about 1 Gyr.

Since the physics of the proposed scenario is rather uncertain, we choose
to model the properties of the halos and their cosmic evolution using a simple
statistical approach.
Through Monte Carlo calculations we take into account the main processes that
play a role in this scenario. These include the rate of cluster-cluster 
mergers in the Universe and their mass ratios, and the fraction of the 
energy dissipated during these mergers that is channelled into 
MHD turbulence and acceleration
of high energy particles (Cassano \& Brunetti 2005; Cassano et al. 2006a).
We refer the reader to these papers for details, 
here we briefly report the essential steps that enter into the calculations : 

\begin{itemize}
\item[{\it i)}]
the formation and evolution of galaxy clusters is computed through 
the extended Press \& Schechter approach (1974, hearafter PS; 
Lacey \& Cole 1993) that is based on the hierarchical theory of 
cluster formation. The PS mass function shows a good agreement with that
derived from N-body simulations, at least for relatively low redshifts
and masses $\sim10^{14}-10^{15}\,h^{-1}\,M_{\odot}$ (\eg Springel et al. 2005),
although it has the tendency to underestimate the number density
of systems with mass $\geq 10^{15}\,h^{-1}\,M_{\odot}$ (\eg Governato et al. 1999; Bode et al. 2001; Jenkins et al. 2001). Given a present day mass and temperature of the parent clusters, the cluster merger history ({\it merger trees}) is obtained making use of Monte Carlo simulations. We simulate the formation history of $\sim 1000$ galaxy clusters with present-day masses in the range $2\cdot 10^{14}-6\cdot 10^{15}\,M_{\odot}$. This allows a statistical description of the cosmological evolution of galaxy clusters and of
the merging events with cosmic time;

\item[{\it ii)}]
the generation of the turbulence in the ICM is estimated for every merger
identified in the {\it merger trees}. The resulting turbulence is assumed to be generated and then dissipated within a time-scale of the order of the cluster-cluster crossing time in that merger\footnote{The cascading time scale of large scale turbulence is expected to be of the same order of cluster-cluster crossing time (\eg Cassano \& Brunetti 2005; Brunetti \& Lazarian 2007).}.
Furthermore, it is assumed that turbulence is generated in the volume swept by the subcluster infalling into the main cluster and that a fraction, $\eta_t$, 
of the $PdV$ work done by this subcluster goes into the excitation of {\it fast magneto--acoustic waves}. The $PdV$ work is estimated as $\approx \rho v_i^2 \pi r_s^2 R_v$,
where $\rho$ is the ICM density of the main cluster averaged over the swept cylinder, $v_i$ is the impact velocity of the two subclusters, $r_s$ is the stripping radius (see also Sect.2.1), and $R_v$ is the virial radius of the main cluster (see Cassano \& Brunetti 2005 for details).

\item[{\it iii)}]
the resulting spectrum of MHD turbulence generated by the chain
of mergers in any
synthetic cluster and its evolution with cosmic time
is computed by taking into account the injection of waves 
and their damping in a collisionless plasma.
Acceleration of particles by this turbulence and their evolution
is computed in connection with the evolution of synthetic clusters 
by solving Fokker-Planck equations and including the relevant 
energy losses.
%\end{itemize}

%\noindent
\item[{\it iv)}]
This procedure allows for the exploration of the statistical properties of radio halos. Following Cassano et al.~(2006), we consider homogeneous models, \ie without spatial variation of the turbulent energy, acceleration rate and magnetic field in the halo volume. We assume a value of the magnetic field, averaged over a region of radius $R_H\simeq 500\,h_{50}^{-1}$ kpc, which scales with the virial mass of clusters, $M_v$ as:

\begin{equation}
<B> \, = B_{<M>} \big({{M_v}\over{<M>}} \big)^{b}
\label{b}
\end{equation}

\noindent
where $b>0$ is a parameter that enters in the model calculations. Eq.~\ref{b} is motivated by numerical cosmological (MHD) simulations that found a scaling of the
magnetic field with temperature or mass of the simulated
clusters (\eg Dolag et al.~2002)\footnote{Dolag et al. (2002) found a scaling $B\propto T^2$, that would imply $B\propto M^{4/3}$ if the virial scaling $M\propto T^{3/2}$ is assumed.}.
\end{itemize}

\begin{figure}
\centerline{
\includegraphics[width=7cm,height=7cm]{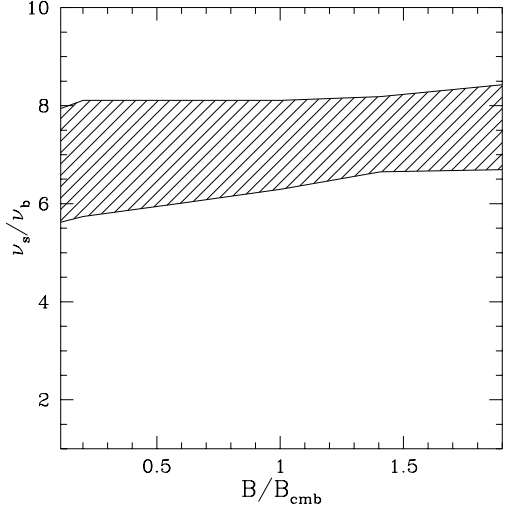}}
\caption{Ratio $\nu_s/\nu_b$ as a function of $B/B_{cmb}$ calculated
with homogeneous models. Particle acceleration is calculated via
TTD resonance with magneto acoustic waves (Brunetti \& Lazarian 2007). 
The shadowed region marks the range of $\nu_s/\nu_b$ obtained assuming 
different energy density of magneto acoustic turbulence,
3\% -- 30\% of the thermal energy density, and duration of
the acceleration process, $\Delta t =$ 2--4 $\times \tau_{acc}$,
where the acceleration time-scale is $\tau_{acc} = \chi^{-1} =
p^2/(4 D_{pp})$.}
\label{fig:ango}
\end{figure}

\subsection{Occurrence of radio halos in galaxy clusters}

\begin{figure*}
\begin{center}
\includegraphics[width=0.4\textwidth]{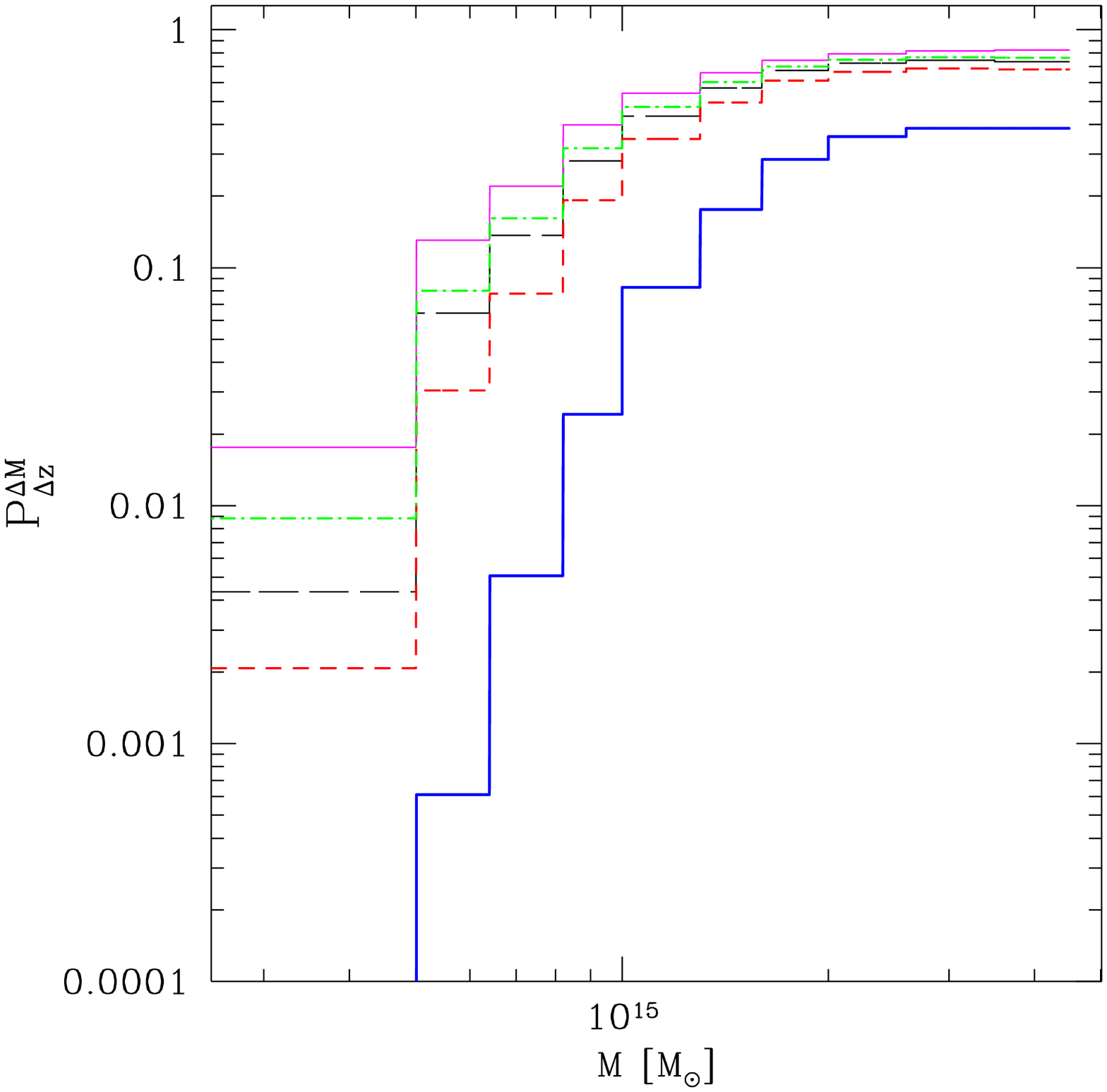}
\includegraphics[width=0.4\textwidth]{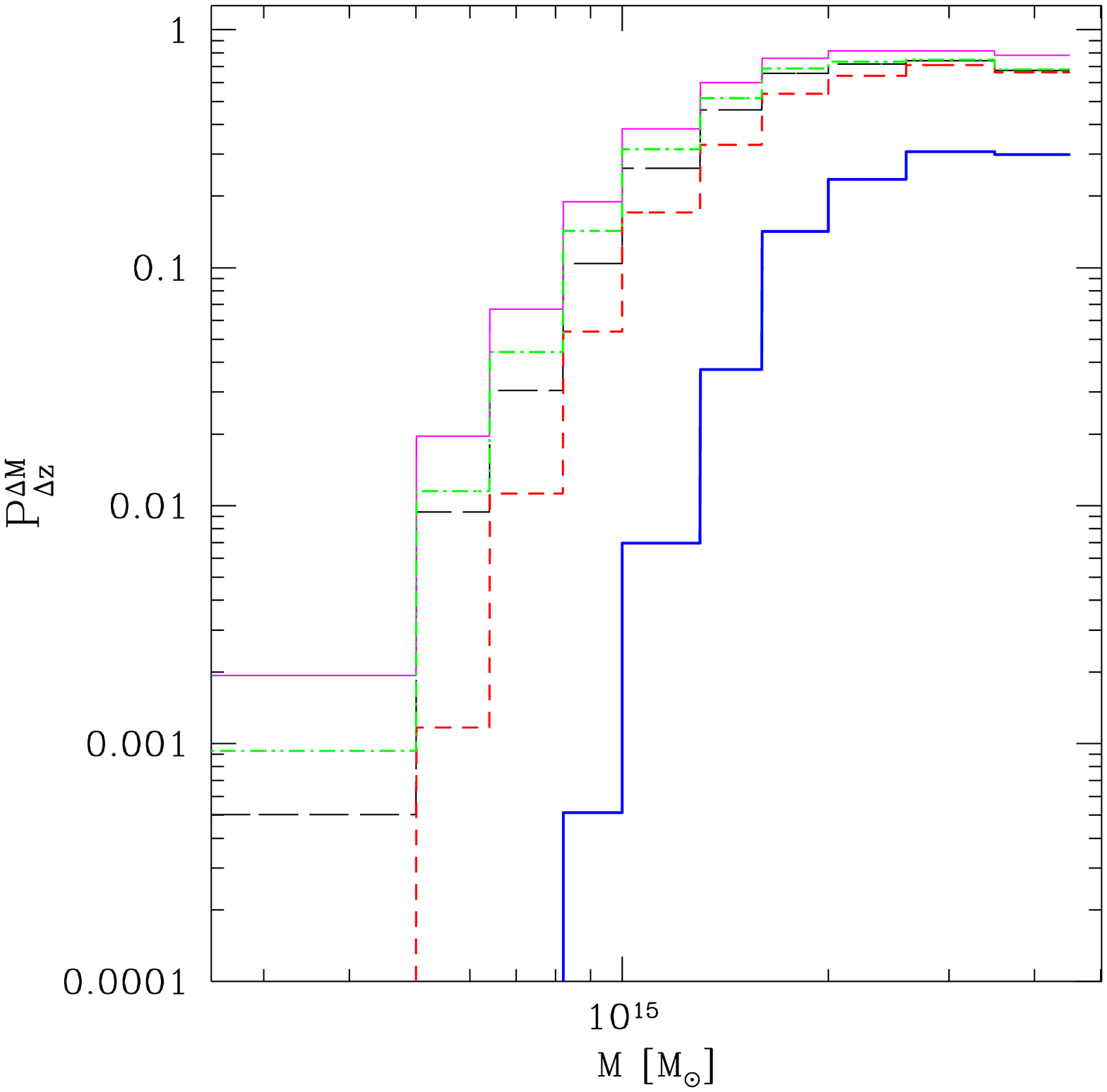}
\caption[]{Fraction of clusters with radio halos, with $\nu_s \geq \nu_o$,
as a function of the cluster mass in the redshift range $0-0.1$ (left panel) 
and $0.4-0.5$ (right panel).
Calculations assume $\nu_o$ = 1.4 GHz, 
240 MHz, 150 MHz, 120 MHz and 74 MHz (from bottom to top).}
\label{Fig.fraction_RH}
\end{center}
\end{figure*} 

Stochastic particle acceleration by MHD turbulence is believed
to be rather inefficient in the ICM. Consequently electrons can 
be accelerated only up to energies of $m_e c^2 \gamma_{max} \leq$ several GeV, 
since at higher energies the radiation losses are efficient and hence dominate. 
The resulting gradual steepening in the theoretical synchrotron spectrum of radio halos 
at high frequencies is consistent with the observed spectral shapes (or with the very steep spectra) of several well studied halos (\eg Schlickeiser et al 1987; Thierbach et al. 2003; Brunetti et al. 2008; Dallacasa et al. 2009).

This steepening will make it difficult to detect these sources
at frequencies larger than the frequency, $\nu_s$, at which the 
steepening becomes severe (see Fig.\ref{fig:ussrh}). $\nu_s$ is expected 
to be a few times larger than the break frequency, $\nu_b$.
$\nu_b$ depends
on the acceleration efficiency
in the ICM, $\chi$, and is defined as (\eg Cassano et al.~2006) :

\begin{equation}
\nu_b \propto \, <B> \gamma_{max}^2 \propto {{<B> \chi^2 }\over{
\big( <B>^2 + B_{cmb}^2 \big)^2 }}
\label{nub}
\end{equation}

\noindent
The {\it Transit Time Damping} (TTD) is the most important collisionless resonance between the magnetosonic waves and particles, and is due to the interaction of the compressible component of the magnetic field of these waves with the particles  (\eg Schlickeiser \& Miller 1998; Cassano \& Brunetti 2005; Brunetti \& Lazarian 2007). In this case $\chi \simeq 4 D_{\rm pp}/p^{2}$, where $p$ is the momentum of the electrons and $D_{pp}$ is 
the electron diffusion coefficient in the momentum space due to the coupling with turbulent waves. Cassano \& Brunetti (2005) derived that in the case of a single merger
between a cluster with mass $M_v$ and a subcluster of mass 
$\Delta M$, $\chi$ can be approximated as:

%\begin{eqnarray}
\begin{equation}
\chi \propto { { \eta_t}\over{ R_H^3 }}
\Big({ {M_{v} +\Delta M}\over{R_v}}\Big)^{3/2}
\frac{r_s^2}{\sqrt{k_B T}}
\times
\Big\{
\begin{array}{ll}
1 & {\rm if}\, r_s \leq R_H \\
(R_H/r_s)^2 & {\rm if}\, r_s > R_H
\end{array}
\label{nub2}
\end{equation}
 
\noindent
where $r_s$ is the stripping radius of the subcluster crossing the main cluster, \ie the distance from the center of the subcluster where the static pressure equals the ram pressure (see Cassano \& Brunetti 2005 for details); $R_H$ is the size of the radio halo, and $R_v$ and $T$ are the virial radius and temperature of the main cluster, respectively.

\noindent
Combined with Eq.~\ref{nub} this implies that larger values of 
$\nu_b$ are expected in the more massive clusters,
$\nu_b \propto (M_v/R_v)^3/T \propto M_v^{4/3}$ (here considering 
for simplicity
a fixed value of $B$, see Cassano et al.~2006 for more general
discussion), and in connection with major merger events, 
$\nu_b \propto (1+\Delta M/M)^3$ (also $r_s$ in Eq.3 increases with $\Delta M/M$).

\noindent
Monte Carlo simulations can now be used to follow cluster-mergers and to explore how different mergers contribute to the acceleration (efficiency) 
of relativistic particles in the ICM. Consequently this allows for a 
statistical modeling of $\nu_b$ within a synthetic cluster sample 
and the derivation of its statistical dependence on cosmic time and 
cluster mass.

\begin{figure*}
\begin{center}
\includegraphics[width=0.4\textwidth]{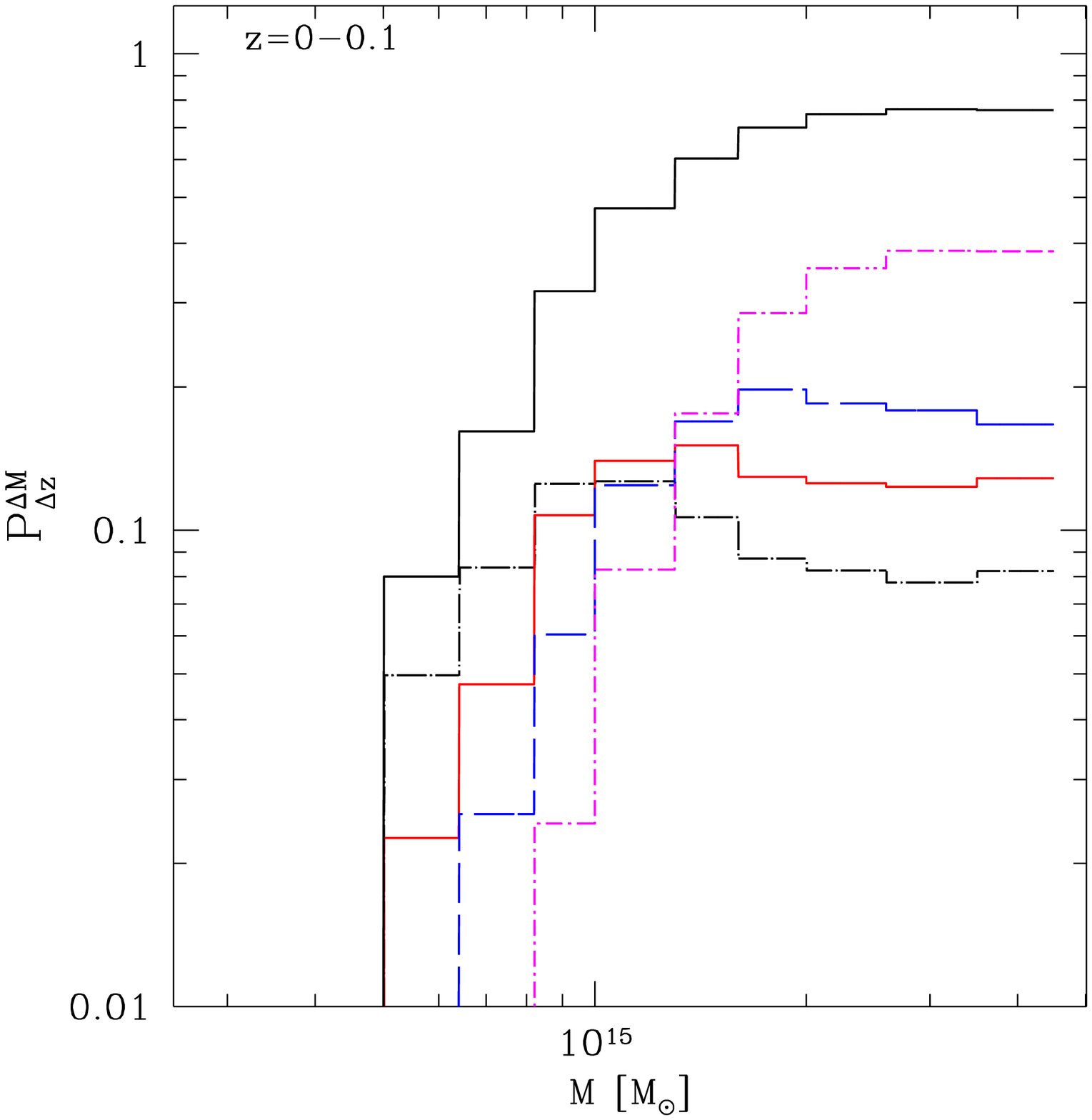}
\includegraphics[width=0.4\textwidth]{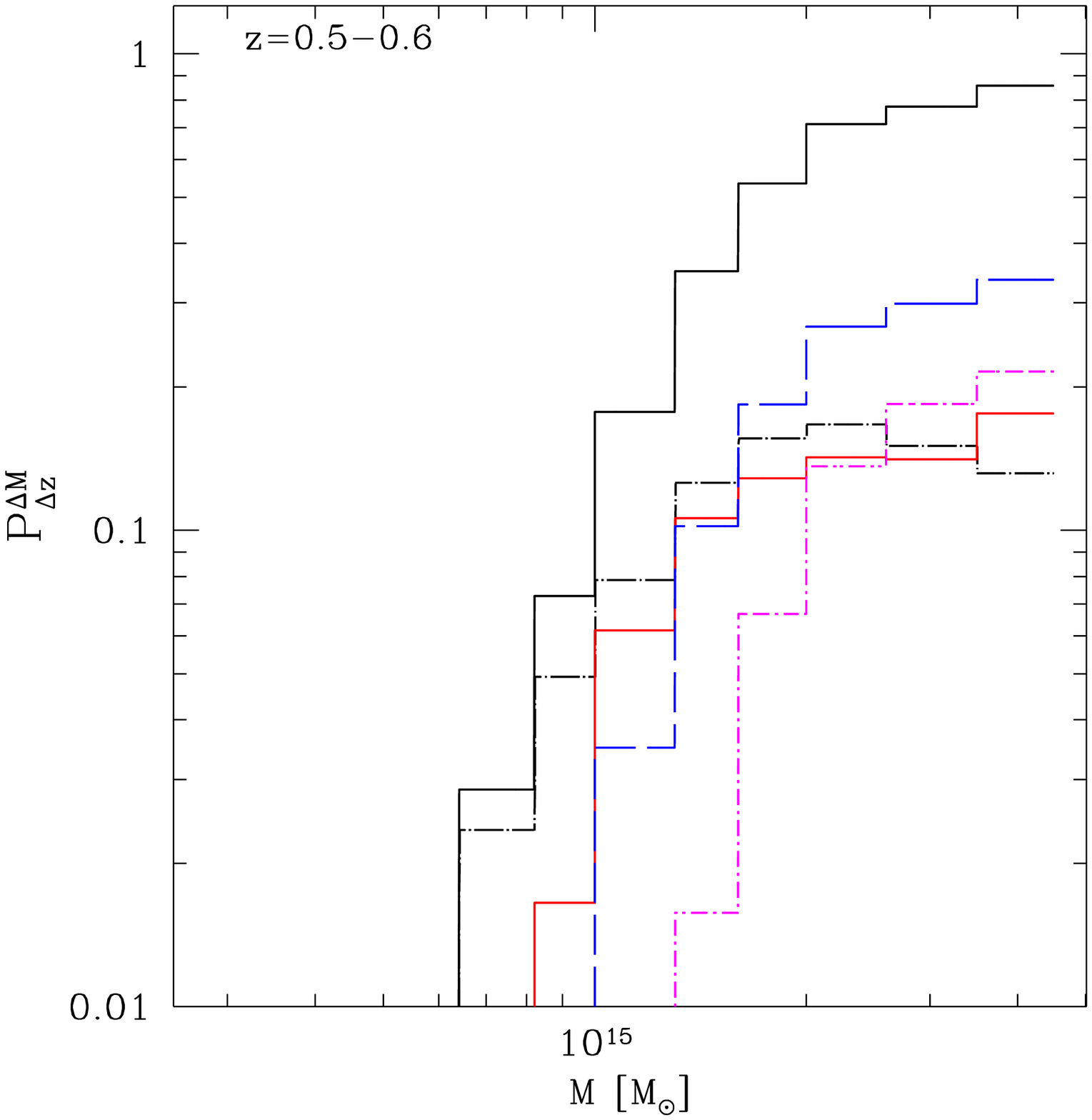}
\caption[]{Fraction of clusters with radio halos 
with $\nu_s \geq$ 120 MHz (black, upper, solid lines) as a function of
the cluster mass in the redshift range $0-0.1$ (left panel) and $0.5-0.6$
(right panel). The fractions of clusters with radio halos $\nu_s$ in
different frequency ranges are also shown : 
$\nu_s \geq 1400$ MHz, $600<\nu_s <1400$ MHz, $240<\nu_s<600$ MHz,
and $120<\nu<240$ MHz (from top to botton).}
\label{Fig.fraction_USSRH}
\end{center}
\end{figure*}

\noindent
Surveys cannot reveal radio halos that have $\nu_s$ smaller
than the observing frequency, since the spectrum of these halos
should be very steep and their emission would fall below the
survey detection limit (Figure 1).
In order to investigate the statistical behavior of the population of radio halos at different frequencies we only take halos to be observable that have $\nu_s \geq \nu_o$.
Figure \ref{fig:ango} shows the ratio $\nu_s/\nu_b$ calculated for 
homogeneous models of radio halos, defining $\nu_s$ as the
frequency where the synchrotron spectrum of these
halos is $\alpha = 1.9$
($\alpha = 1.9$ calculated between $\nu_s/2.5$ and $\nu_s$ to mimic
600-1400 MHz spectra); as $\nu_s/\nu_b$ is only mildly dependent on magnetic field strength and the assumed fraction of turbulent energy injected, we adopt $\nu_s \sim 7 \nu_b$. A statistical modeling of $\nu_b$ provides a statistical evaluation of $\nu_s$ in the synthetic cluster sample.

\noindent
In the context of the turbulent acceleration model for giant radio
halos energetics arguments imply that halos with $\nu_s \geq$ 1 GHz 
must be generated in connection with the most energetic merger-events 
in the Universe. Only these mergers may allow for the efficient acceleration 
necessary to have relativistic electrons emitting at these 
frequencies (Cassano \& Brunetti 2005). 
Present surveys carried out at $\nu_o \sim$ 1 GHz 
detect radio halos only in the most massive and merging
clusters (\eg Buote 2001, Venturi et al. 2008), and their 
occurrence has been used to constrain the value of 
$\eta_t \approx 0.1-0.3$ in the models (Cassano \& Brunetti 2005).
Similar energetics arguments can be used to
claim that radio halos with 
smaller values of $\nu_s$ must be more common, since they can be generated
in connection with less energetic phenomena, \eg major mergers between
less massive systems or minor mergers in massive systems (\eg
Eqs.~\ref{nub}-\ref{nub2}), that are more common in the 
Universe (\eg Cassano et al.~2008).

In Fig.\ref{Fig.fraction_RH} we plot the fraction of clusters that host radio halos with $\nu_s\geq\nu_o$ as a function of the cluster mass and considering
two redshift ranges : $0-0.1$ (left panel) and $0.4-0.5$ (right panel);
this is obtained assuming a reference set of model parameters,
namely: $<B>=1.9\, \mu$G, $b=1.5$, $\eta_t=0.2$ (see also Cassano et al.~2006).
As expected, the fraction of clusters with halos increases at smaller 
values of $\nu_o$, and the amount of this increment depends on the 
considered mass and redshift of the parent clusters, being larger at 
smaller cluster masses and at higher redshifts.

Fig.\ref{Fig.fraction_USSRH} we plot the fraction 
of radio halos with $\nu_s \geq$ 120 MHz (black upper line) 
and the differential contribution to this 
fraction from radio halos with $\nu_s$ in four frequency 
ranges (see figure caption for details).
For nearby systems (Fig.\ref{Fig.fraction_USSRH}, {\it Left Panel}), a 
significant fraction of massive clusters, $M_v > 10^{15}\,M_{\odot}$,
is expected to host radio halos with $\nu_s \geq$ 120 MHz; a sizeable 
fraction of them with $\nu_s > 600$ MHz (blue and magenta lines). 
On the other hand, the majority of radio halos in clusters with 
mass $M_v\ltsim 10^{15}\,M_{\odot}$ would have very steep spectra
if observed at GHz frequencies, $\nu_s < 600$ MHz (red line and 
black dot-dashed line).
Our calculations suggest that a similar situation is expected for
clusters at higher redshift (Fig.\ref{Fig.fraction_USSRH}, 
{\it Right Panel}).
Radio halos with larger values of $\nu_s$ become much rarer with increasing
redshift, mainly because 
the unavoidable inverse Compton losses at these redshifts
limit the maximum energy of the accelerated electrons in these
systems. At $z > 0.5$,
only merging clusters with mass $M_v\gtsim
2 \cdot 10^{15}\,M_{\odot}$ have a sizeable chance to host giant 
radio halos with $\nu_s \geq 1.4$ GHz, and an increasing contribution
to the percentage of radio halos at higher redshift comes from
halos with lower $\nu_s$.

\subsection{The radio halo luminosity function}

The luminosity functions of radio halos (RHLFs), \ie the number of halos
per comoving volume and radio power, with $\nu_s \geq 1.4$ GHz has been 
derived by Cassano et al. (2006a) as:

\begin{equation}
{dN_{H}(z)\over{dV\,dP(1.4)}}=
{dN_{H}(z)\over{dM\,dV}}\bigg/ {dP(1.4)\over dM}
\label{RHLF}
\end{equation}

\noindent
where $dN_{H}(z)/dM\,dV$ is the theoretical mass function of radio halos with 
$\nu_s \geq 1.4$ GHz, that is obtained combining Monte Carlo calculations of the fraction of clusters with halos and the PS mass function of clusters (\eg Eq.~18 in Cassano et al.~2006).

$dP(1.4)/dM$ can be estimated from the correlation between
the 1.4 GHz radio power, $P(1.4)$, and the mass of the
parent clusters, that is observed for radio halos 
(\eg Govoni et al. 2001; Cassano et al. 2006a). 
Cassano et al.~(2006) discussed the $P(1.4)$--$M_v$ correlation 
in the context of the turbulent acceleration model and have shown that
the slope is consistent with the observed value ($\alpha_M=2.9\pm0.4$) 
for a well constrained region of parameter space ($B_{<M>}$, $b$, 
and $\eta_t$; Figure 7 in Cassano et al.~2006); model parameters 
adopted in the present paper, $<B>=1.9\, \mu$G, $b=1.5$, $\eta_t=0.2$, 
fall in this range. In particular, the value of the derivative $dP(1.4)/dM$ in
Eq.~\ref{RHLF} depends on the set of paramenters ($B_{<M>}$, $b$) that, in the 
case of the reference model we are using in this paper, sets $\alpha_M = 3.3$.

To derive the RHLF at frequency $\nu_o$ the
contribution of all radio halos with $\nu_s \geq \nu_o$ should be
taken into account.
We first obtained the RHLF for halos with $\nu_s$ in
a given frequency interval, \eg ${\Delta \nu_s}_i$, and then combine 
the different contributions from the considered
intervals ${\Delta \nu_s}_i$ :

\begin{equation}
{dN_{H}(z)\over{dV\,dP(\nu_o)}}=
\sum_i 
\big( {dN_{H}(z)\over{dM\,dV}} \big)_{{\Delta \nu_s}_i}
\big( {dP(\nu_o)\over dM} \big)_{{\Delta \nu_s}_i}^{-1}
\label{RHLF_nuo}
\end{equation}

To derive the contribution to the RHLF from radio halos
with $\nu_s \geq$ 1.4 GHz we should calculate $dP(\nu_o)/dM$ 
for these halos. This can be estimated from the $P(1.4)-M_v$ correlation assuming a monochromatic radio power of these halos at $\nu_o$ :

\begin{equation}
P_{1.4}(\nu_o, M_v) = P_{1.4}(1.4, M_v)
\Big(\frac{1400\,\mathrm{MHz}}{\nu_o}\Big)^{\alpha}\,,
\label{scale1}
\end{equation}

\noindent
where $P_{1.4}(1.4, M_v)$ is the monochromatic radio power at 1.4 GHz
from the $P(1.4)-M_v$ correlation, and 
$\alpha\sim 1.3$ the typical spectral index of these halos, $P(\nu)\propto\nu^{-\alpha}$ 
(\eg Ferrari et al.~2008).

\noindent
We now consider the case of halos with $\nu_s <$ 1.4 GHz.
The bolometric synchrotron power of radio halos is expected to 
scale with $\nu_b$ and $B$ (\eg Cassano et al.~2006) :

\begin{equation}
P_{syn} \approx 
P(\nu_b)\nu_b \propto B \nu_b \Rightarrow P(\nu_b)\propto B\,.
\label{psyn}
\end{equation}

\noindent 
From Eqs.\ref{nub}-\ref{nub2} it is clear that clusters with the same
mass $M_v$ (and magnetic field $B$) at redshift $z$ could have
different values of $\nu_b$, depending on the merger event responsible
of the generation of the radio halo.
Yet, Eq.~\ref{psyn} implies that for a fixed cluster mass
(and consequently for a fixed value of the magnetic field)
the synchrotron radio power emitted at the break frequency, $P(\nu_b)$, 
is constant.
In addition, homogeneous models, that consider an average value
of $B$ and $\gamma_{max}$ in the halo volume,
also imply that $P(\nu_b) \nu_b \propto 
P(\nu_s) \nu_s$ (Fig.~\ref{fig:ango}).
From Eq.~\ref{psyn}, we can then derive the monochromatic 
radio power at $\nu_o$ 
of halos with a given $\nu_s$ :

\begin{equation}
P_{\nu_s}(\nu_o, M_v)=
P_{\nu_s}(\nu_s, M_v)\Big(\frac{\nu_s}{\nu_o}\Big)^{\alpha}
=
P_{1.4}(1.4, M_v)
\Big(\frac{\nu_s}{\nu_o}\Big)^{\alpha}
\label{scale2}
\end{equation}

\noindent
This allows the evaluation of $(dP(\nu_o)/dM)_{{\Delta \nu_s}_i}$ 
starting from $dP(1.4)/dM$, and thus to derive Eq.~\ref{RHLF_nuo}.

\noindent
We also note that from Eqs.~\ref{scale1} and \ref{scale2} one 
has :

\begin{equation}
P_{\nu_s}(\nu_o, M_v)=
P_{1.4}(\nu_o, M_v) \Big(\frac{\nu_s}{1400\,\mathrm{MHz}}\Big)^{\alpha}
\label{scale3}
\end{equation}

\noindent
i.e. radio halos with synchrotron spectra steepening at smaller 
frequencies will also have monochromatic radio powers at $\nu_o$
smaller than those of radio halos with larger $\nu_s$.

\begin{figure}
\begin{center}
\includegraphics[width=0.5\textwidth]{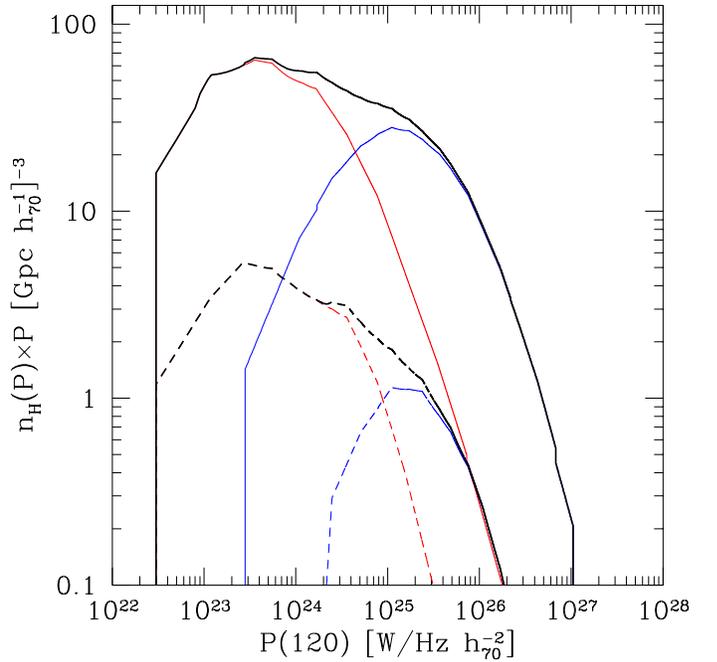}
\caption[]{Radio halo luminosity function at $\nu_o$=120 MHz (black
lines) for clusters 
at redshift $0-0.1$ (solid thick lines) and $0.5-0.6$ (dashed thin lines). 
The contribution from halos with $120<\nu_s < 600$ MHz (red lines) and 
with $\nu_s \geq 600$ MHz (blue lines) are also shown.}
\label{Fig.RHLF}
\end{center}
\end{figure} 

\begin{figure}
%\centerline{
\includegraphics[width=0.5\textwidth]{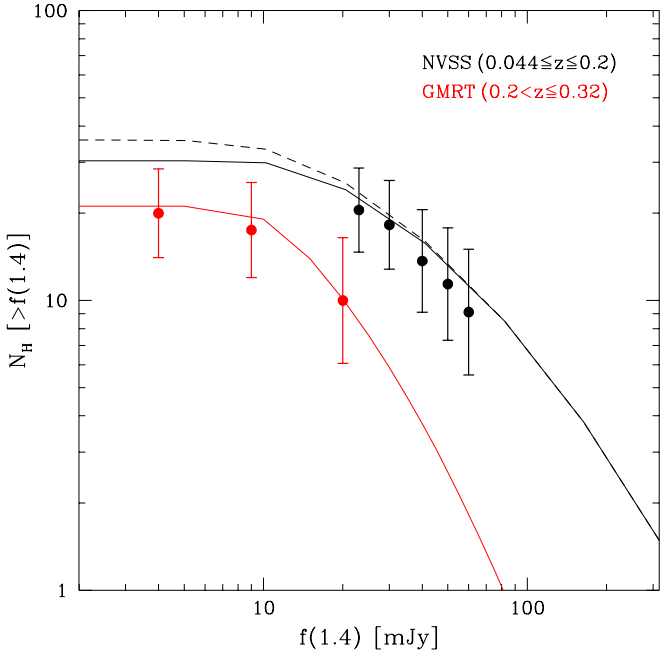}
%}
\caption{
All sky integrated RHNCs for: (1) z=0.044-0.2, obtained by considering a
minimum mass of clusters constrained at any $z$ by the
XBACS X-ray flux limit (Ebeling et al. 1996) (dashed black line),
and by combining the above mass--constraint with that implied by 
the NVSS sensitivity (following Cassano et al.~2008, 
see their Figure 3) (solid black line);
(2) z=0.2-0.32, obtained by considering the X-ray luminosity-range of the
GMRT cluster sample (Venturi et al.~2007, 2008).
Black points are the observed RHNC of giant radio halos from NVSS selected
clusters in the redshift range 0.044-0.2 and re-normalized to account
for the NVSS and XBACS sky coverage (and XBACS completeness).
Red points are the observed RHNC of giant radio halos in the GMRT cluster
sample (with redshift z=0.2-0.32) and re-normalized to account for the 
sky coverage of the GMRT cluster sample.}
\label{fig:NC_NVSS_GMRT}
\end{figure}

\noindent 
As a relevant example, 
in Fig.\ref{Fig.RHLF} we report the expected RHLF at 120 MHz 
(black lines) for $z=0-0.1$ (solid thick lines) and $z=0.5-0.6$ 
(dashed thick lines), where we also show the relative contribution 
from 
halos with $\nu_s<600$ MHz (red lines) and $\nu_s>600$ MHz (blue lines). 

\noindent
As already discussed in Cassano et al.~(2006), the shape of
the RHLF flattens at lower radio 
powers due to the decrease of the efficiency of particle 
acceleration in less massive clusters. 
We note that halos with $\nu_s > 600$ MHz (blue lines,
Fig.~\ref{Fig.RHLF}) do not contribute to lower radio powers in the RHLF.
This is because higher-frequency halos are generated in very energetic
merger events, and must be extremely rare in smaller systems
and consequently their monochromatic radio power is larger
than that of halos with $\nu_s < 600$ MHz (red lines, Fig.~\ref{Fig.RHLF}).
Finally, we note that with increasing redshift the 
RHLFs decrease due to the evolution of the cluster mass function with $z$ and to the 
evolution with $z$ of the fraction of galaxy clusters with radio halos 
(Fig.~\ref{Fig.fraction_RH}, see also Cassano et al.~2006). 
The evolution of the RHLF with $z$ is stronger at higher radio powers, 
where the dominant contribution to the RHLF comes from halos 
with larger $\nu_s$ and the fraction of clusters hosting these halos 
decreases more rapidly with redshift (\eg Fig.~\ref{Fig.fraction_USSRH}).

\section{Number Counts of Radio Halos and LOFAR surveys at 120 MHz}

\noindent
It has been shown that model expectations of the occurrence of radio halos 
observed at $\nu_o$= 1.4 GHz are consistent with the fraction of radio 
halos with cluster mass (Cassano et al.~2008) and with the number counts 
of nearby radio halos (Cassano et al.~2006).

\noindent
As already discussed, in this paper we adopt a reference model 
with parameters: $<B>=1.9\, \mu$G, $b=1.5$, $\eta_t=0.2$. 
In Fig.~\ref{fig:NC_NVSS_GMRT} we report number counts of giant radio halos expected with these parameters compared with radio halo counts from the NVSS survey at low redshift, $0.044\leq z\leq 0.2$ (Giovannini et al.~1999) and from the {\it GMRT Radio Halo Survey} at intermediate redshift, $z=0.2-0.32$ (Venturi et al.~2007, 2008). The latter is a pointed survey down to $70\,\mu$Jy/beam at 610 MHz of a sample of $\sim 50$ galaxy clusters extracted from the REFLEX (B\"ohringer et al. 2004) and eBCS (Ebeling et al. 1998, 2000) cluster catalogs. The clusters have $z=0.2-0.4$ and $L_{X[0.1-2.4\,kev]}\geq 5\cdot 10^{44}$ erg/s (the X-ray sample is complete for $z\leq0.32$; see Cassano et al. 2008). All halos in the survey have $1.4$ GHz follow up.
Beside the fair agreement between expectations and observations
(see caption), we note that the GMRT radio halo survey is sufficiently sensitive to catch relatively faint halos and to constrain the flattening of the distribution of number counts of radio halos (RHNC) at lower fluxes.

\noindent
Encouraged by these results, in this Section we derive expected RHNC at 120 MHz and explore the potential of the upcoming LOFAR surveys.

\noindent
Because in our simplified procedure the radio power of halos scales 
with a spectral slope $\alpha = 1.3$ (Eqs.~\ref{scale1}--\ref{scale3})
and the great majority of halos is at $z \sim$0.2--0.4,
in the following we neglect K-correction.\footnote{For simplicity 
we also consider observables those halos with $\nu_s \approx \nu_o$
regardless of their redshift. This would slightly affect only the
number counts of halos with $\nu_s \leq \nu_o (1+z)$ that represent
a minimal fraction of our halo population.}

\subsection{LOFAR Surveys}

LOFAR will carry out surveys between 15 MHz and 210 MHz with
unprecedented sensitivity and spatial resolution 
(\eg R\"ottgering et al. 2006).
Also, the unprecedented (u,v) coverage of LOFAR at short baselines
maximizes the instrument capability to detect extended sources with 
low surface brightness such as radio halos.
These surveys will constrain models for the
origin of diffuse radio emission in galaxy clusters.
In this paper we assume an observing frequency $\nu_o$= 120 MHz where
LOFAR will carry out the deepest large area radio surveys 
(\eg R\"ottgering et al. 2006).

\begin{figure*}
\begin{center}
\includegraphics[width=0.4\textwidth]{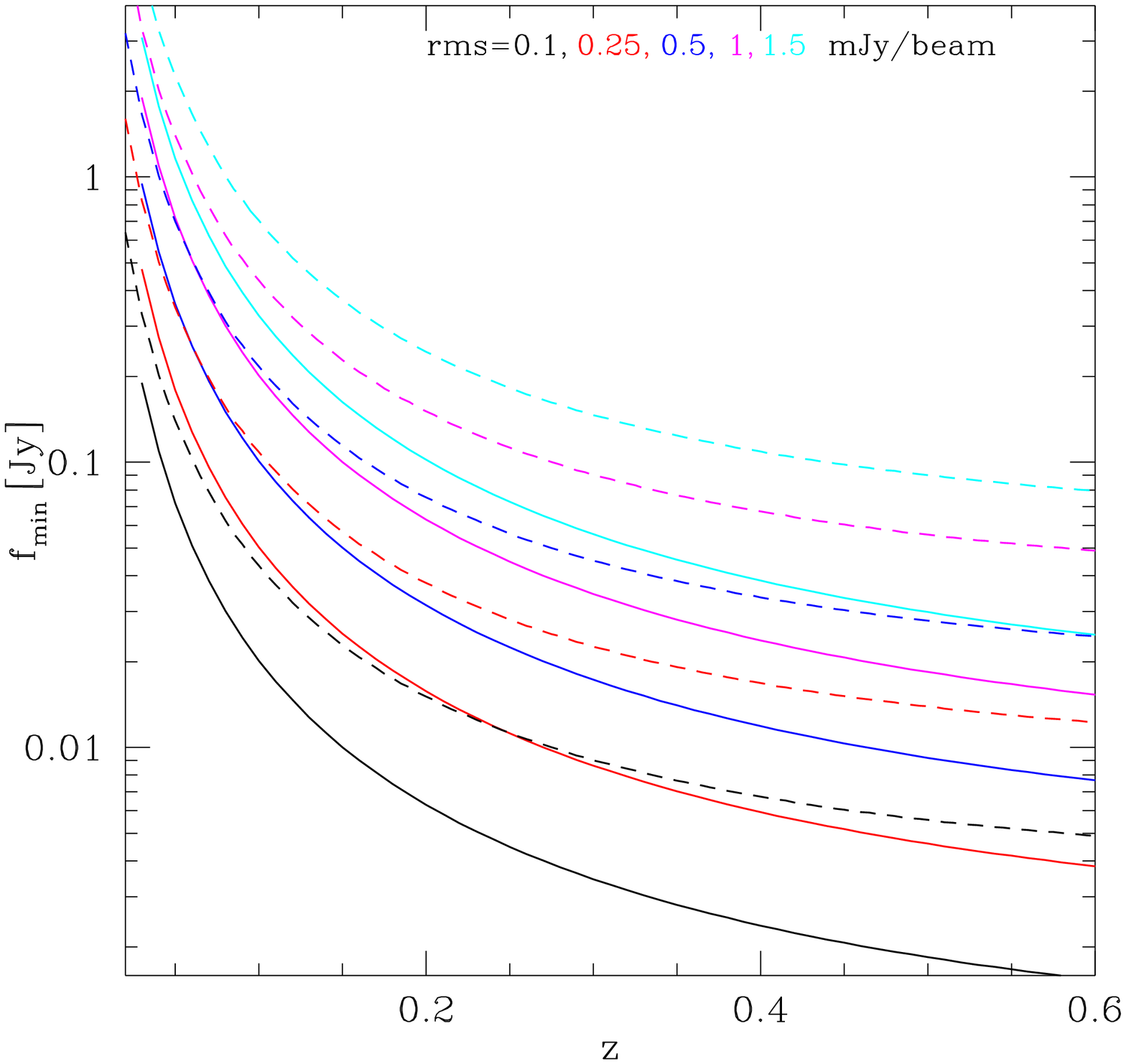}
\includegraphics[width=0.4\textwidth]{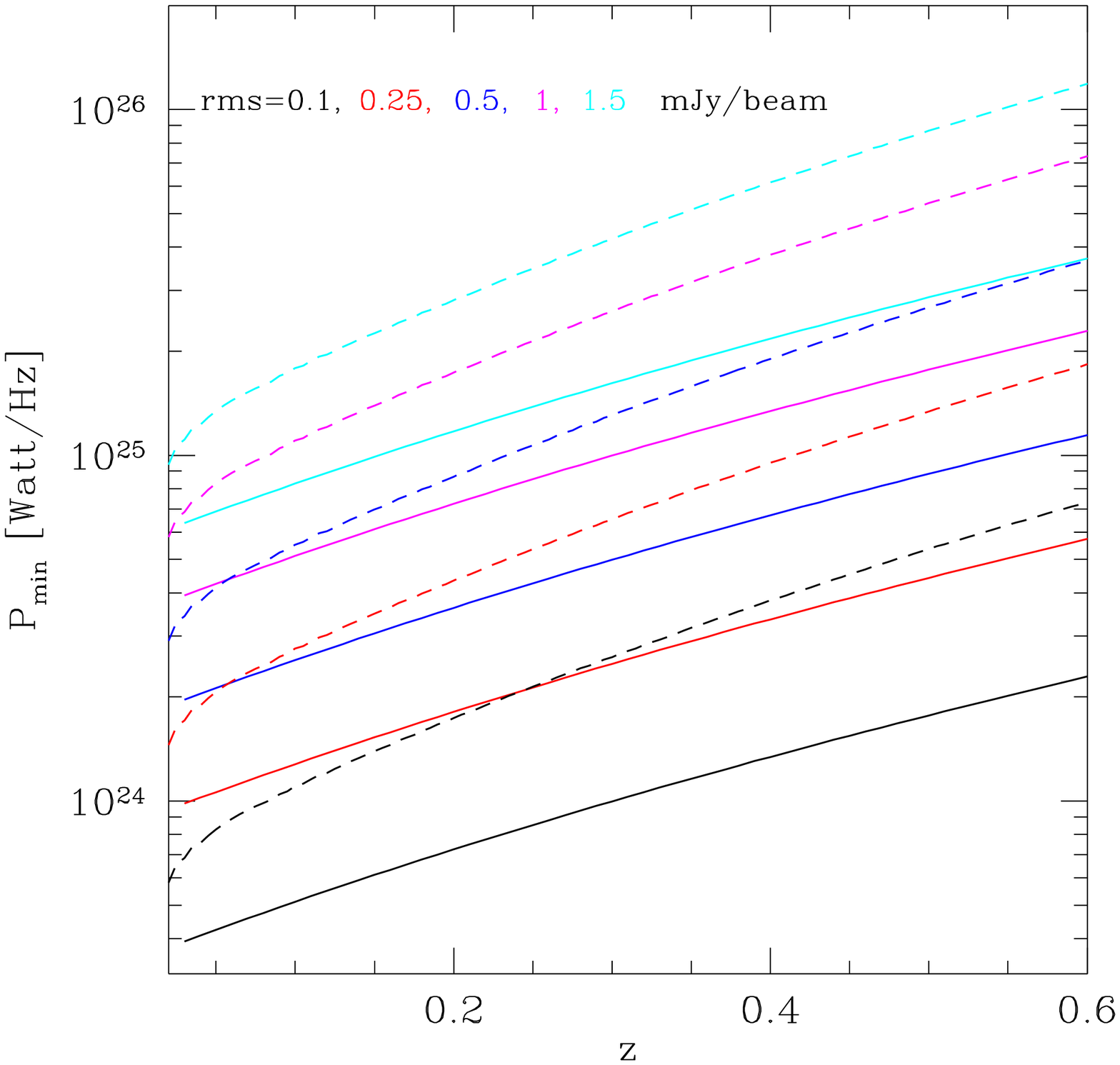}
\caption[]{
Minimum flux (left panel) and power (right panel) of detectable radio
halos at 120 MHz using Eq.~\ref{fmin} (solid lines) and Eq.~\ref{fmin2} 
(dashed lines).
$\xi \, F$ = 0.1, 0.25, 0.5, 1.0, 1.5 mJy/b are assumed
(from bottom to top), and beam=25$\times$25 arcsec.}
\label{Fig.flim_plim}
\end{center}
\end{figure*}

\noindent
The crucial step in our analysis is the estimate of the minimum
diffuse flux from giant radio halos (integrated over a scale 
of $\sim$ 1 Mpc) that is detectable by these surveys
as a function of redshift.
This depends on the brightness profiles of radio halos
that is known to smoothly decrease with distance from the cluster center
(\eg Govoni et al. 2001). Consequently the outermost, low brightness, 
regions of halos will be difficult to detect.

\noindent However what is important is the capability to detect the central, 
brightest, regions of radio halos in the survey images. Following Brunetti et al.~(2007), 
we consider a shape of the radial profile of radio halos that is obtained from the analysis of well studied halos. We assume a circular observing beam = 25$\times$25 arcsec, and follow two complementary approaches\footnote{The 120 MHz LOFAR survey will have a full
resolution of 5--6 arcsec, thus we are considering the case of tapered
images that increase the sensitivity to extended emission without changing significantly the point source sensitivity (due to the large number of inner LOFAR stations).}:

\begin{itemize}

\item[{\it i)}]
since radio halos emit about half of their total radio flux 
within their half radius (Brunetti et al. 2007), 
we estimate the minimum flux of a detectable halo, $f_{min}(z)$, 
through the requirement that the mean brightness within 
$R_H /2$, $\frac{f_{min}/2}{\pi (\theta_H/2)^2}$, is $\xi$ times the $rms$, $F$, of the survey, i.e. :

\begin{figure*}
\begin{center}
\includegraphics[width=0.4\textwidth]{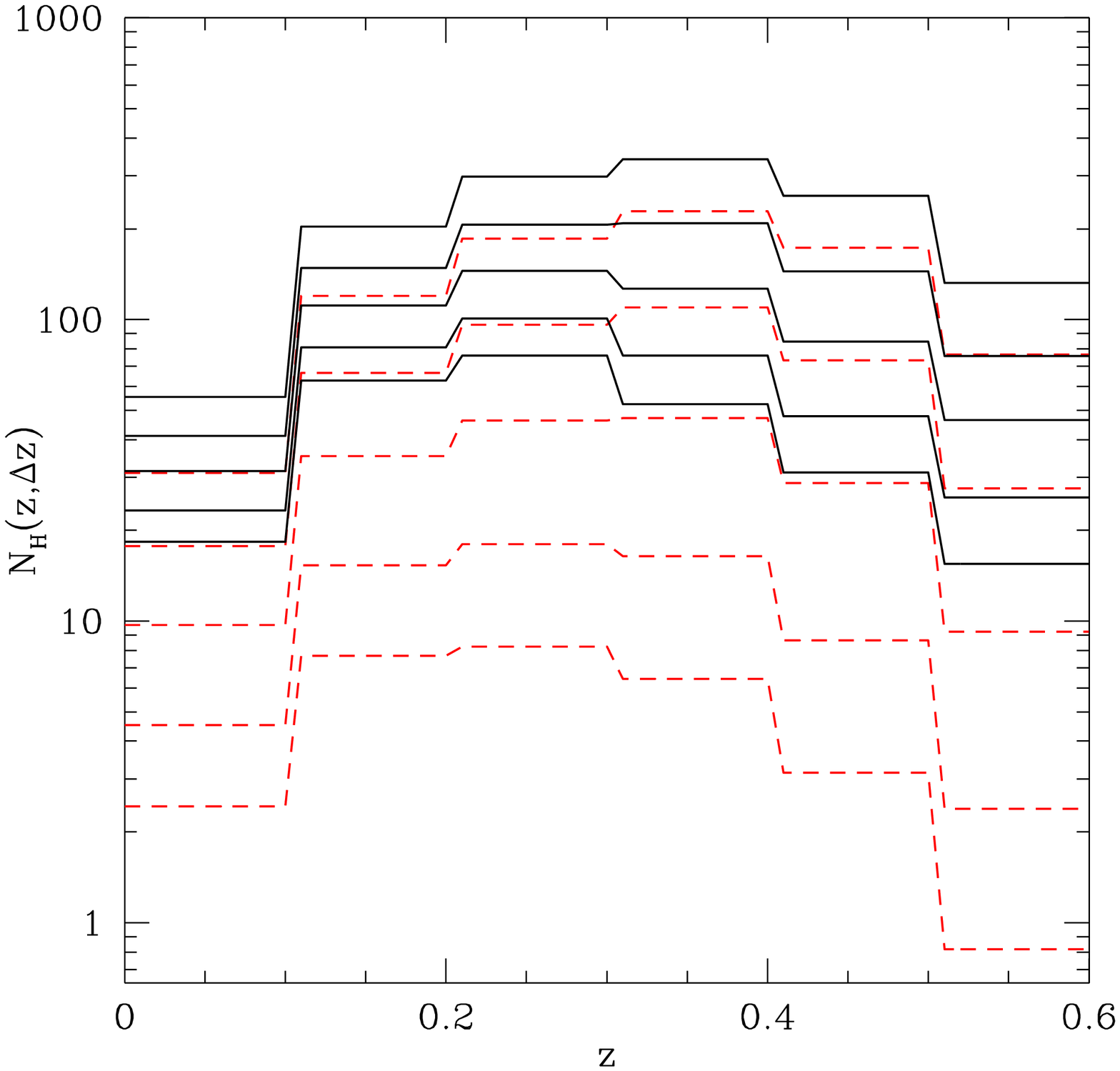}
\includegraphics[width=0.4\textwidth]{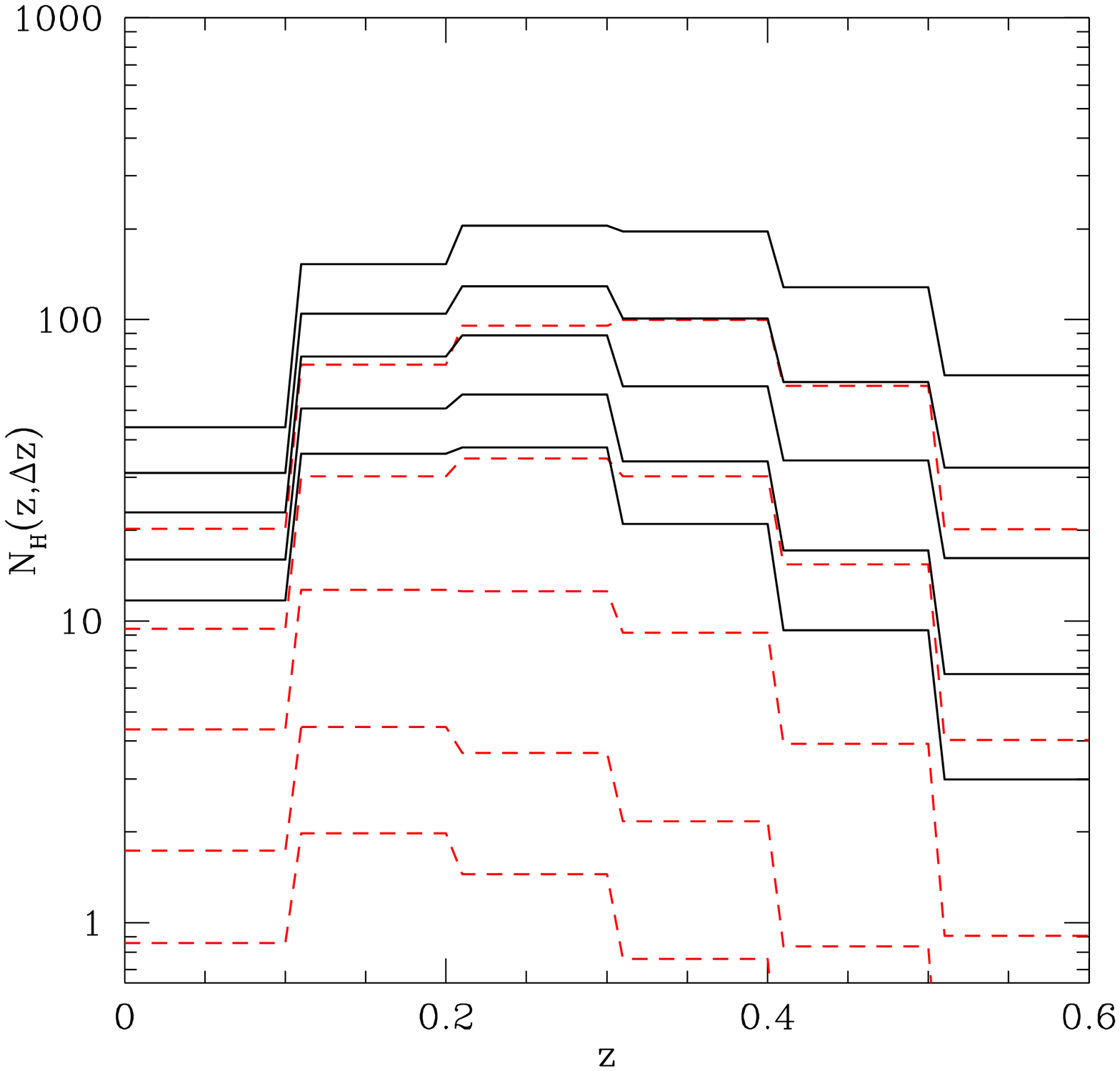}
\caption[]{
Number (all-sky) of radio halos with $\nu_s \geq$ 120 MHz (black lines)  
as a function of redshift that can be expected at the sensitivity of 
LOFAR surveys. 
Calculations are performed following approaches {\it i)} in Sect.~2.1
(left) and {\it ii)} (right), assuming $\xi \, rms$=
0.1, 0.25, 0.5, 1.0, 1.5 (bottom to top).
Red lines give the number counts of radio halos with 
$120 \leq\nu_s\leq 600$ MHz.}
\label{Fig.RHNC}
\end{center}
\end{figure*} 

\begin{equation}
f_{min}(z)\simeq 10^{-3}\Big(\frac{\xi \, F}{0.5 \mathrm{mJy/beam}}\Big)\,
\theta_{H}^2(z)\,\, [\mathrm{mJy}]
\label{fmin}
\end{equation}

\noindent
where $\theta_{H}(z)$ is the angular size of radio halos, in arcseconds, at a given
redshift; allowing for the detection of diffuse halo emission in the images 
produced by the survey.
Injection of fake radio halos in the (u,v) plane of interferometric
data from NVSS observations show that radio halos at z $\leq$ 0.3 become 
visible in the images as soon as their flux approaches that obtained 
by Eq.~\ref{fmin} with $\xi \sim 1-2$ (Cassano et al.~2008);

\item[{\it ii)}]
following a second approach we estimate the minimum flux of a detectable halo
through the requirement that the average brightness within 5 observing beams 
is $3 \times \xi$ times the $rms$, $F$, of the survey. The minimum flux is
obtained through the condition :

\begin{equation}
2 \pi \int_0^{b_s} I(b) b db = 5 \, S_{beam} \, ( 3 \, \xi \, F)
\label{fmin2}
\end{equation}

\noindent
where $I(b)$ is the typical radial profile of halos (Brunetti et al.~2007),
$S_{beam}$ is the beam area, and $b_s = (5\, S_{beam}/\pi)^{1/2}$.
The aim of this second approach is to avoid any bias due to the 
redshift of the halos and is motivated by the fact that in the first approach 
the sensitivity limit is reached in a fairly large area (many beams) 
for nearby radio halos, but only in a few beam area in the case of halos 
at z=0.5-0.6.
\end{itemize}

Fig.\ref{Fig.flim_plim} shows $f_{min}$ of radio halos as a function of
redshift (left panel), and the corresponding minimum radio power (right
panel), obtained following the two approaches and assuming 
$\xi \,F$= $0.1 \, {\ldots} 1.5$ mJy/beam (see Figure caption).

\begin{figure}
\begin{center}
\includegraphics[width=0.4\textwidth]{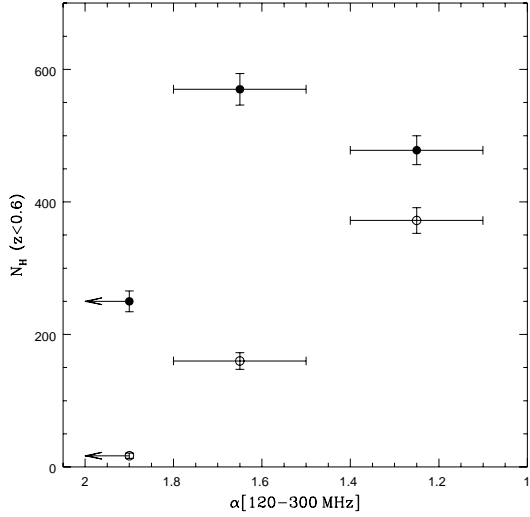}
\caption[]{Spectral index distribution of radio halos, calculated
between 120-330 MHz, for the number counts in Fig.~\ref{Fig.RHNC} (left panel)
and assuming $\xi \, F$= 0.1 (filled symbols) and 0.5 mJy/b (empty
symbols). Spectral indices are calculated for homogeneous models,
with parameters given in Sect.~2, assuming the $\nu_b$ distribution
derived from the Monte Carlo--synthetic clusters. Radio halos are binned according to the following values of $\nu_b$: $\nu_b>90$ MHz, $35<\nu_b<90$ MHz, and $\nu_b<35$ MHz
(from right to left). The horizontal bars encompasse the variation in $\alpha$ due to the range of $\nu_s/\nu_b$ in Fig.~\ref{fig:ango}.}
\label{Fig.alpha}
\end{center}
\end{figure} 

\noindent 
Given the RHLF ($dN_H(z)/dP(\nu_o)dV$) the number counts of radio halos 
with $f\geq f_{min}(z)$ in a redshift interval, $\Delta z=z_2-z_1$, 
is given by :

\begin{equation}
N_{H}^{\Delta_z}(>f_{min}(z))=\int_{z=z_1}^{z=z_2}dz' ({{dV}\over{dz'}})
\int_{P_{min}(f_{min}^{*},z')}{{dN_H(P(\nu),z')}\over{dP(\nu_o)\,dV}}
dP(\nu_o)  
\label{RHNC}
\end{equation}

\noindent 
In Fig.~\ref{Fig.RHNC} we show the all-sky number 
of radio halos with $\nu_s \geq$ 120 MHz in different redshift 
intervals detectable 
by typical LOFAR surveys with different sensitivities 
($0.1$ {\ldots} $1.5$ mJy/beam, see figure caption) and following
the approach {\it i)} (Left Panel) and {\it ii)} (Right Panel). 

\noindent
The LOFAR all sky survey (\eg R\"ottgering 2009, priv. comm.) is expected 
to reach an rms=0.1 mJy/beam at 120 MHz. Considering the 
case {\it i)} (Fig.~\ref{Fig.RHNC}, Left Panel) 
with $\xi \sim 2-3$, we predict that this survey will detect more than 
350 radio halos at redshift $\leq$0.6, in the northern 
hemisphere ($\delta\geq0$) and at high Galactic latitudes ($|b|\geq 20$).
This will increase the statistics of radio halos by about a factor of $20$ with
respect to that produced with the NVSS.  
The LOFAR commissioning MS$^3$ 
survey is expected to reach sensitivities $\approx$0.5 mJy/b at 150 MHz. 
Based on our results, $\approx$100 radio halos are expected to be 
discovered through this
survey within one year time-scale.

\begin{figure}
\begin{center}
\includegraphics[width=0.4\textwidth]{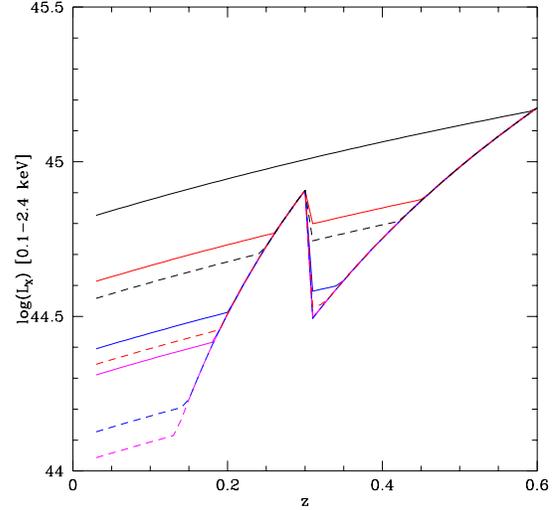}
\caption[]{Minimum X-ray luminosity of clusters with detectable
halos at $\nu_o$=120 MHz.
Calculations are shown for $\xi \, F$=0.25 (dashed lines) and
1 mJy/b (solid lines), and for different ranges of $\nu_s$ : 
$120<\nu_s<240$ MHz (black lines), $240<\nu_s<600$ MHz (red lines),
$600<\nu_s<1400$ MHz (blue lines) and $\nu_s \geq 1.4$ GHz (magenta lines)
(from top to bottom).}
\label{Fig.popopo}
\end{center}
\end{figure}

The spectral properties of the population of radio halos
visible by the future radio surveys at low frequencies are expected
to change with increasing the sensitivity of these surveys.
In Fig.~\ref{Fig.RHNC} 
we show the total number of halos with $\nu_s \geq$ 120 MHz (solid lines)
together with the number of halos with a spectral steepening at low
frequencies, $120 \leq \nu_s \leq 600$ MHz.
The latter class of radio halos has a synchrotron spectral
index $\alpha > 1.9$ between 250-600 MHz, and would 
become visible only at low frequencies, $\nu_o < 600$ MHz.
We find that about 55\% of radio halos in the LOFAR all sky survey
at 120 MHz is expected to belong to this class of ultra-steep spectrum
radio halos, while radio halos with larger $\nu_s$ are expected to
dominate the population in shallower surveys.
This is simply due to the fact that, for the reasons explained
in Sect.~2.2, low frequency 
radio halos are expected to populate the low power-end of the
RHLF (e.g. Fig.~\ref{Fig.RHLF}).
Complementary information is given
in Fig.~\ref{Fig.alpha} that shows the expected distribution of halo 
spectral indices, 
that refer to the number distributions in Fig.~\ref{Fig.RHNC}, and
its evolution with sensitivity of radio observations;
spectra in Fig.~\ref{Fig.alpha} are calculated between 120--300 MHz 
assuming homogeneous models.

\noindent
Ultra-steep spectrum 
halos are a unique prediction of the turbulent re-acceleration
model (\eg Brunetti et al. 2008) and our expectations demonstrate the potential of LOFAR in
constraining present models for the origin of radio halos.

\begin{figure*}
\begin{center}
\includegraphics[width=0.42\textwidth]{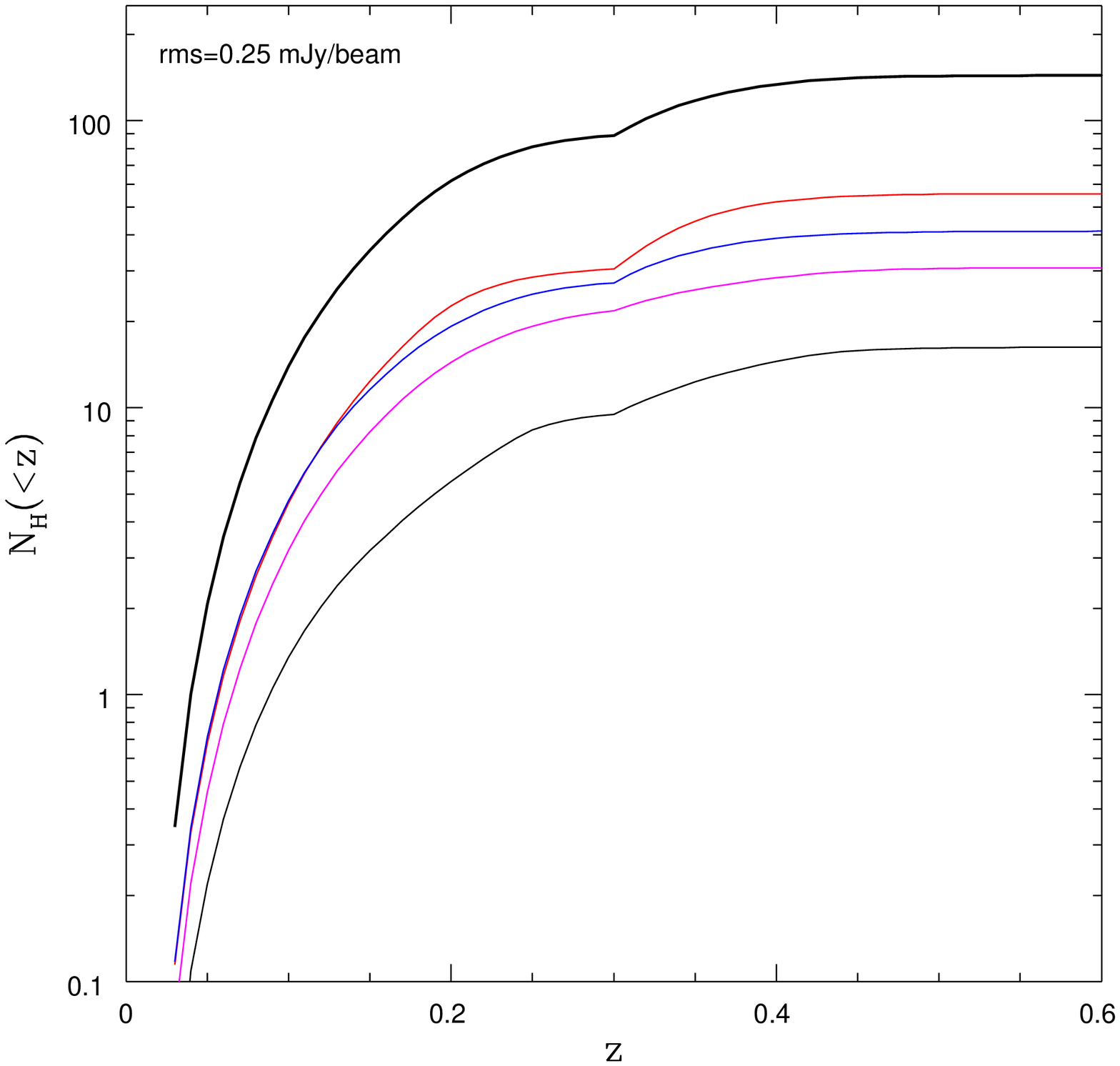}
\includegraphics[width=0.42\textwidth]{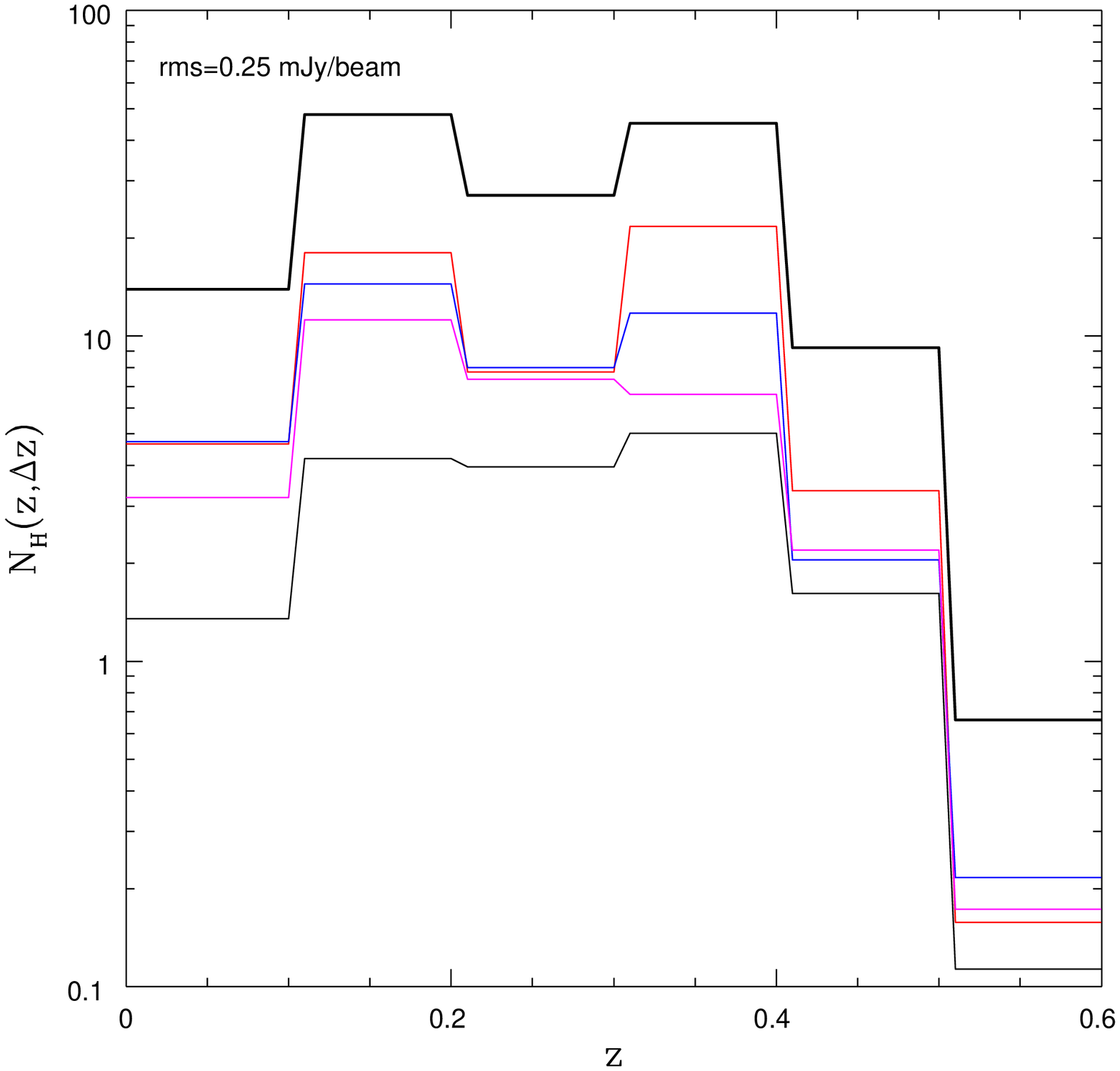}
\includegraphics[width=0.42\textwidth]{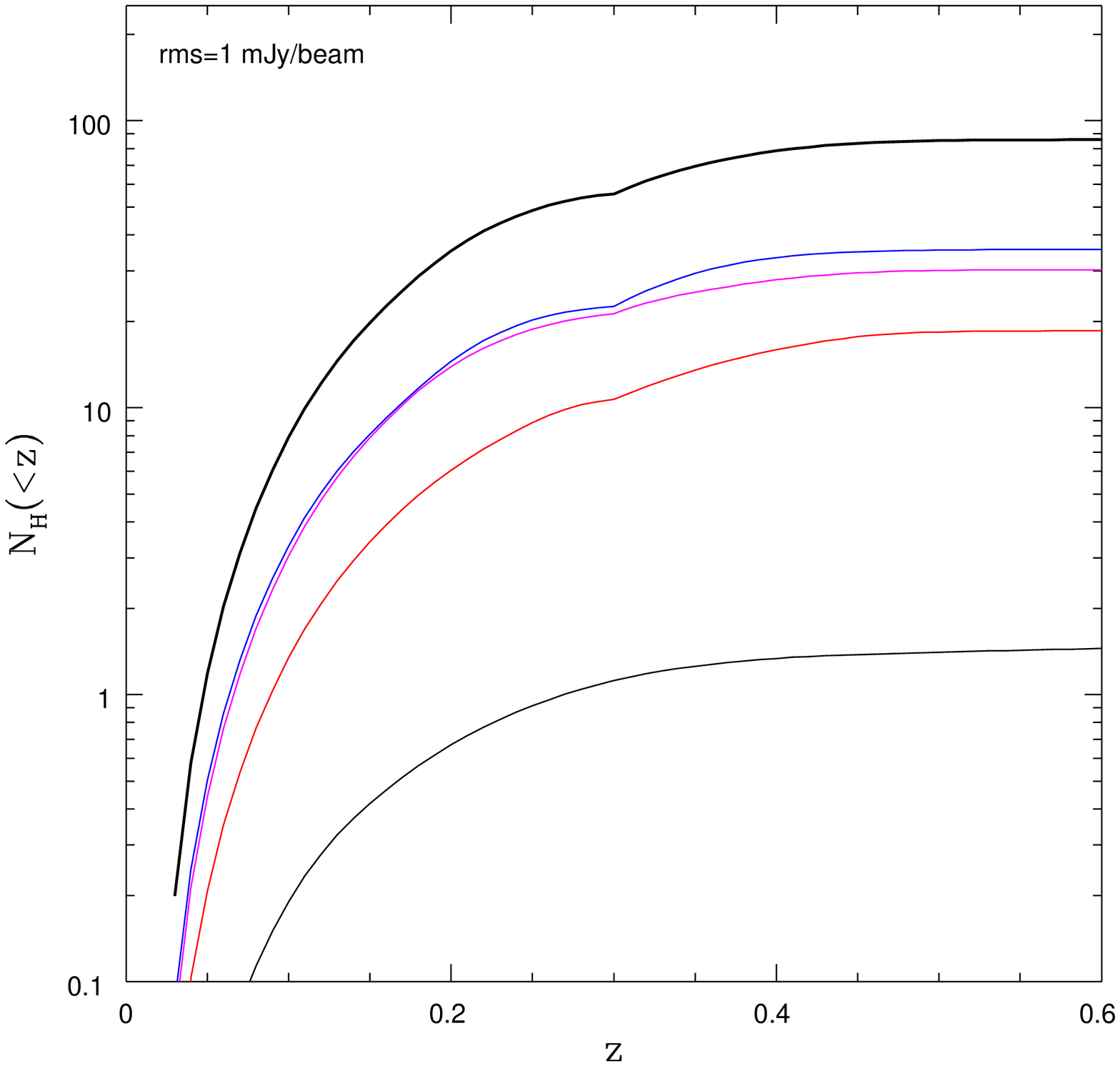}
\includegraphics[width=0.42\textwidth]{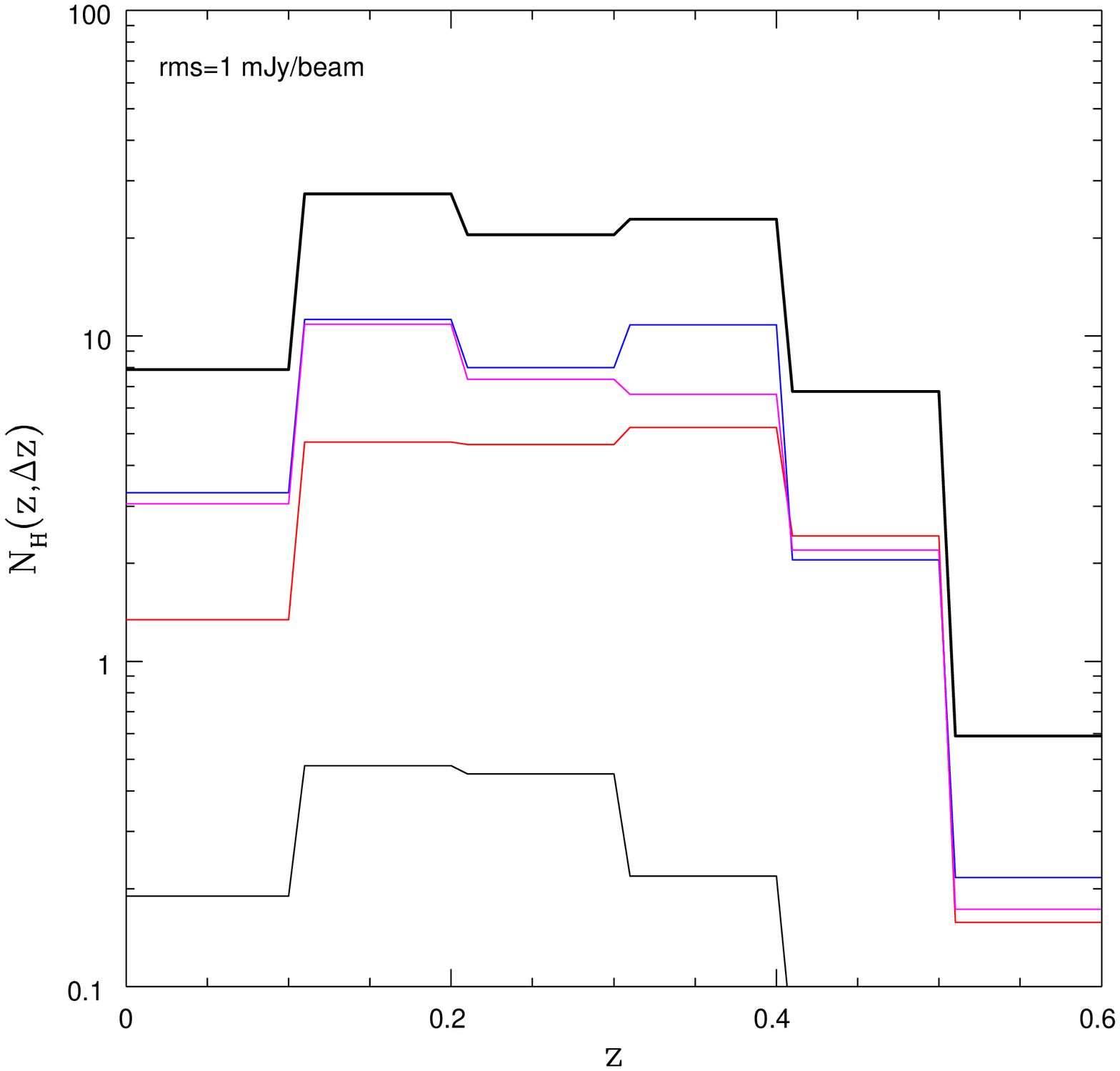}
\caption[]{Integrated (left) and differential (right) number counts of 
radio halos from radio follow up of eBCS and MACS clusters (see text).
Calculations are shown for $\xi \, F$= 0.25 (upper panels) and
1.0 mJy/b (lower panels) at 120 MHz.
Thick (black) solid lines give the case $\nu_s \geq$ 120 MHz, while 
differential contributions are shown with different colors :
$\nu_s \geq 1.4$ GHz (magenta lines), $600<\nu_s<1400$ MHz (blue lines), 
$240<\nu_s<600$ MHz (red lines) and $120<\nu_s<240$ MHz (black thin lines).}
\label{Fig.Counts_X}
\end{center}
\end{figure*}

\begin{figure*}
\begin{center}
\includegraphics[width=0.42\textwidth]{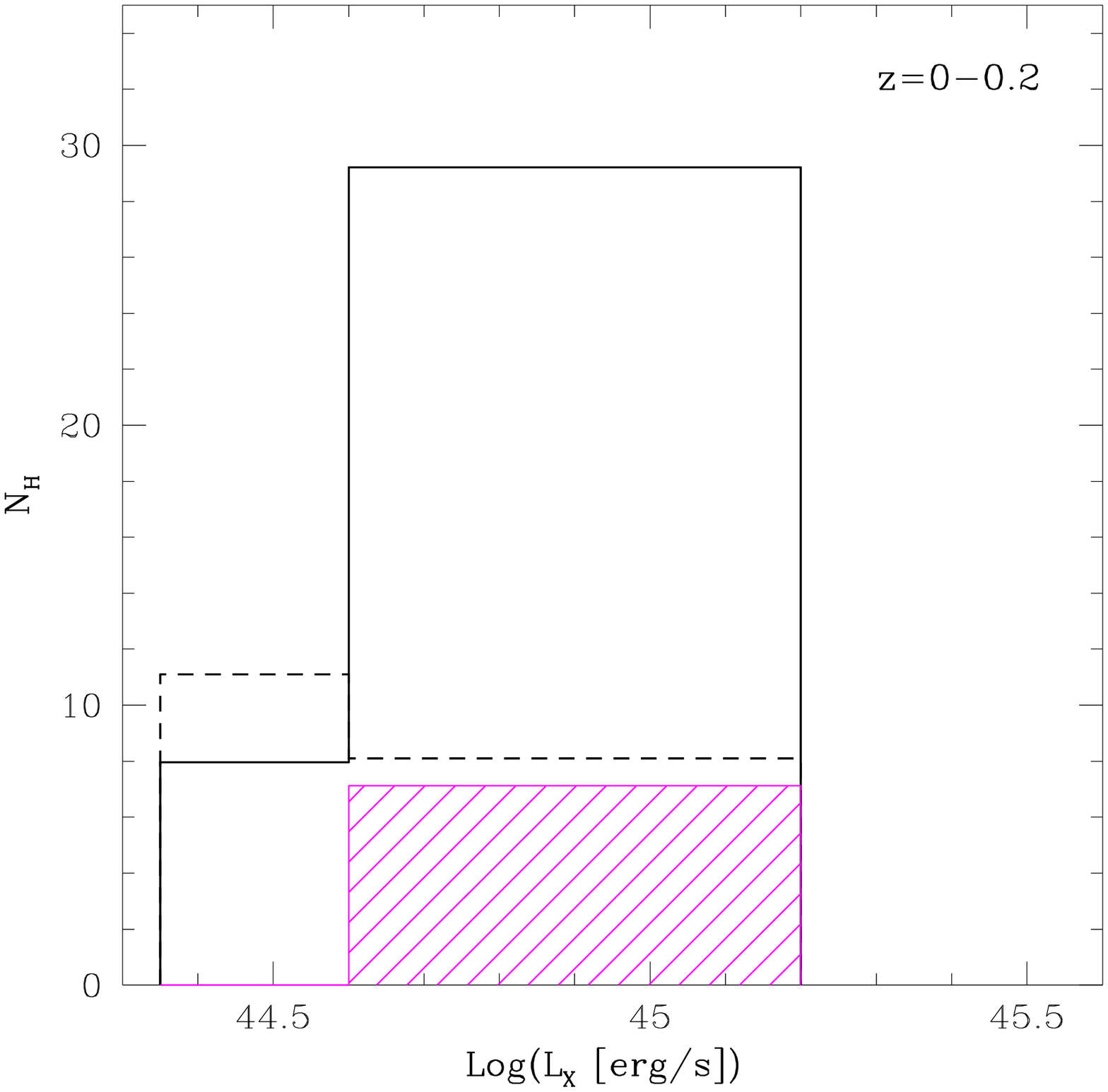}
\includegraphics[width=0.42\textwidth]{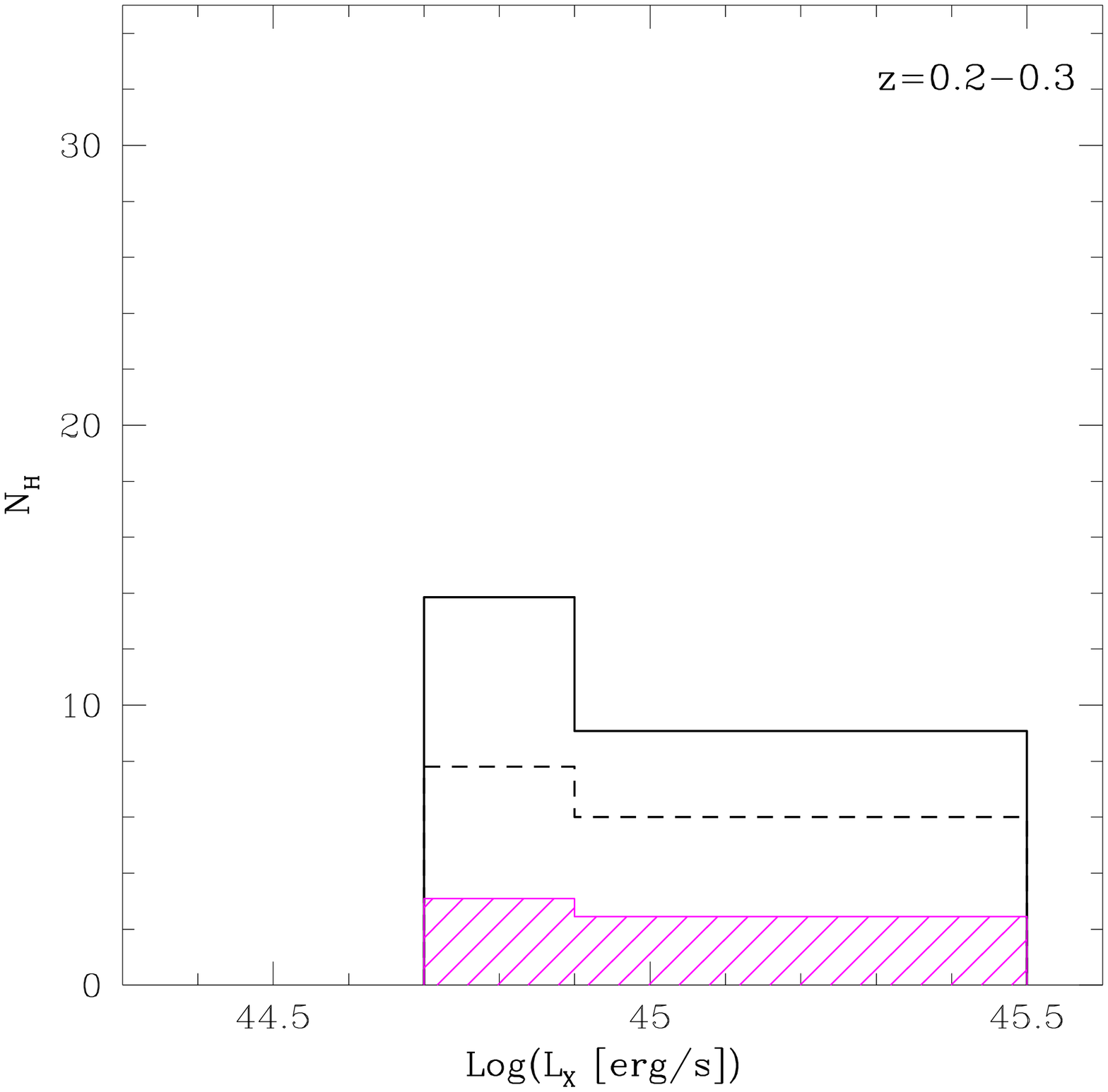}
\caption[]{
Number distributions of radio halos in eBCS clusters from radio
follow up are shown in different X-ray luminosity bins of the hosting
clusters and in two redshift bins : 0--0.2 (left panel) and
0.2--0.3 (right panel).
Calculations are obtained assuming a sensitivity at 120 MHz,
$\xi \, F$= 1 mJy/b and beam = 25$\times$25 arcsec.
Solid lines show the number of radio halos with $\nu_s \geq$ 120 MHz,
while the dashed areas mark the contribution from halos
with $120 < \nu_s < 600$ MHz.
For comparison, 
the dashed lines show the expectations assuming that a fixed fraction of
galaxy clusters = 30\% host radio halos and that all radio halos have
$\nu_s \geq$ 1400 MHz.}
\label{Fig.histo_Lx}
\end{center}
\end{figure*} 

\subsection{Application to X-ray selected cluster samples}

Although unbiased surveys of radio halos provide an important way 
to address the occurrence of these sources (Sect.~3.1), a potential 
problem with these approaches is the identification of radio halos 
and of their hosting clusters. This is because 
radio halos constitute only a very small fraction of the entire 
radio source population and need to be distinguished from confused 
regions of superpositions for radio AGNs and starbust galaxies.

\noindent
Alternatively, an efficient approach is to exploit deep LOFAR surveys 
at low radio frequencies of X-ray selected samples of galaxy clusters.

Here we derive the number of radio halos, and their flux and redshift 
distribution, that are expected from LOFAR observations of sample of X-ray selected clusters.

There are several catalogs of X-ray selected clusters in the northern
hemisphere that contain clusters extracted from the {\it ROSAT} All-Sky Survey
(RASS, Tr\"umper 1993).
At redshift $\leq 0.3$ the {\it ROSAT} Brightest Cluster Sample and 
its extension at lower X-ray fluxes (eBCS, Ebeling et al. 1998, 2000) 
and the Northern {\it ROSAT} All-Sky (NORAS) Cluster Survey 
(B\"ohringer et al. 2000) provide cluster catalogs with X-ray 
flux $f_{X[0.1-2.4]\mathrm{keV}}\gtsim 3\cdot 
10^{-12} \mathrm{erg\, s^{-1} cm^{-2}}$; the eBCS is 75\% complete down 
to this flux limit. The extension towards larger redshifts of these
catalogs is given by the Massive Cluster Survey (MACS, Ebeling 
et al. 2001) that contains 
clusters with $f_{X[0.1-2.4]\mathrm{keV}}\geq 1\cdot 10^{-12} 
\mathrm{erg\, s^{-1} cm^{-2}}$ at z=0.3-0.6.
All these surveys have optical follow-ups and provide a useful 
starting point to pick up radio halos in LOFAR surveys.

A well-known correlation exists between the synchrotron power of giant radio
halos and the X-ray luminosity of the hosting clusters,
$P(1.4) \propto L_X^x$, with $x \simeq 2$ (\eg Liang et al.~2000;
Bacchi et al.~2003; En\ss lin \& R\"ottgering 2002; Cassano et al.~2006;
Brunetti et al.~2009). This implies that the X-ray flux limit of the survey, 
$f_X$, is related to the radio flux of halos. 
The minimum flux of radio halos that can be detected at redshift $z$ 
is given by the maximum value between the minimum radio flux due to 
the sensitivity of radio surveys (Sect.~3.1) and
that constrained by $f_X$ through the radio -- X-ray correlation.
To address this issue at $\nu_o$=120 MHz in 
the case of radio halos with $\nu_s \geq 1.4$ GHz we assume a
correlation between the monochromatic radio luminosity at 120 MHz and
$L_X$ rescaled from that at 1.4 GHz through Eq.~\ref{scale1}. 
For halos with lower $\nu_s$ (yet $\nu_s >$ 120 MHz)
the correlation between the radio luminosity at 120 MHz and $L_X$
is obtained via Eq.~\ref{scale2} that accounts for the smaller radio 
power expected in the case of halos with steeper spectrum (Sect.~2.2).

\noindent 
In this Section we model the sensitivity of LOFAR at 120 MHz 
following the approach {\it i)} in Sect.~3 (Eq.~\ref{fmin}).
More specifically, to detect radio halos, we consider 120 MHz LOFAR 
follow ups of a cluster 
catalog obtained by combining the eBCS (at $z \leq 0.3$) and the 
MACS ($0.3 <z <0.6$) samples, and assume references sensitivities 
of radio observations $\xi\,F=0.25$ and 1 mJy/beam.
The minimum $L_X$ of clusters where these radio observations 
are expected to detect giant radio halos results from the combination of the above
radio sensitivity and the minimum $L_X$ in cluster catalogs at redshift
$z$, and is shown in Fig.~\ref{Fig.popopo} by considering different
$\nu_s$ (see Figure caption for details).
One may note that at intermediate redshift and at higher
redshift the luminosity-limit is driven by the X-ray flux limit of 
the eBCS and MACS catalogs, respectively.
On the other hand, we expect that in the redshift range where the minimum $L_X$ is constrained 
by the radio sensitivity, radio halos with $\nu_s$ in the range 
120--240 MHz (Fig.~\ref{Fig.popopo}, black lines) can be detected 
in clusters with X-ray 
luminosity about 3 times larger than that of clusters with $\nu_s \geq 1.4$ GHz halos (Fig.~\ref{Fig.popopo}, magenta lines).

\noindent 
In Fig.\ref{Fig.Counts_X} we show the cumulative and
differential number counts of radio halos expected from the
LOFAR follow up of eBCS and MACS clusters at 120 MHz.
This is obtained through Eq.~\ref{RHNC} and by taking into
account both the selection 
criteria shown in Fig.\ref{Fig.popopo} and the sky coverage of eBCS
and MACS surveys. The inflection in the number counts at $z=0.3$ is due to the
change in the X-ray selection criteria (see Fig.\ref{Fig.popopo}) moving from the eBCS ($z\leq0.3$) to the MACS ($z\geq0.3$) cluster sample.
We expect that the LOFAR all sky survey, with a planned sensitivity in
line with the case $\xi\,F=0.25$ mJy/beam (Fig.\ref{Fig.Counts_X}, upper 
panels), will discover about 130 radio halos out of the $\sim400$ clusters in the
eBCS and MACS catalogs.
Remarkably, about 40\% of these radio halos are expected with $\nu_s \leq
600$ MHz, thus halos with extreme steep spectra at GHz frequencies.
The majority of radio halos in eBCS and MACS clusters is expected
at z = 0.2-0.4, while the small number of clusters at $z \geq 0.5$ with
X-ray flux above the flux limit of the MACS catalog does not
allow a statistically solid expectations, although we may expect 
a couple of radio halos hosted in MACS clusters at this redshift.
At this redshift we expect that only major mergers in massive clusters ($M_v\geq 2\cdot 10^{15}\,M_{\odot}$) can generate radio halos with $\nu_s\geq 1.4$ GHz (Fig.~\ref{Fig.fraction_USSRH}, right panel). The powerful radio halo recently discovered in the cluster MACS\,J0717.5 +3745 (\eg Bonafede et al. 2009; van Weeren et al. 2009) is consistent with these expectations.

\noindent
Fig.~\ref{Fig.Counts_X} (lower panels) shows the expected number counts of
radio halos assuming the more conservative case $\xi\,F=1$ mJy/b that
is suitable to explore the potential 
of the LOFAR MS$^3$ commissioning survey.
In this case about 80 radio halos are expected to be found in eBCS 
and MACS clusters; about 20 of these halos are expected with $\nu_s <$ 600 MHz.
We notice that the number of expected radio halos from follow up of
eBCS and MACS clusters increases by less than 
a factor of 2 considering a substantial drop in radio sensitivity from 
$\xi\,F=1$ to 0.25 mJy/b. 
This is not surprising as the majority of radio halos that are expected 
to be discovered
by deep radio observations should be found in galaxy clusters with 
X-ray luminosity below the luminosity-threshold of the eBCS and MACS 
catalogs (\eg Fig.~\ref{Fig.popopo}). 

The eBCS cluster sample contains 300 galaxy clusters at $z<0.3$ and covers the northern hemisphere. The redshift and X-ray luminosity distribution of
eBCS clusters is public (Ebeling et al. 1998, 2000) and thus we can 
provide a more quantitative expectation based 
on \eg the more conservative case,
MS$^3$--like that assumes $\xi\,F=1$ mJy/beam at 120 MHz
(in this case the selection function of
clusters in the $L_X$-$z$ plane is reported in Fig.~\ref{Fig.popopo}, 
solid lines at $z<0.3$).

\noindent
In Fig.\ref{Fig.histo_Lx} we show the distribution of the expected 
radio halos in the eBCS clusters in two redshift 
intervals: $0-0.2$ and $0.2-0.3$ (left and right panels, respectively). 
We find that radio observations at 120 MHz are expected to
discover radio halos in about $60$ clusters, \ie in about 20\% 
of eBCS clusters. In addition,
about $12$ of these halos are expected to have very 
steep radio spectra, $\nu_s<600$ MHz (magenta shadowed 
region in Fig.~\ref{Fig.histo_Lx}).
 
\noindent
Finally, the percentage of clusters with radio halos is expected to increase with 
the X-ray luminosity of the hosting clusters.
This is particularly relevant in the redshift interval 
$z=0-0.2$ when compared to expectations calculated under the
assumption that the fraction of clusters hosting radio halos is
constant with cluster mass (Fig.\ref{Fig.histo_Lx} dashed lines, 
see caption). 
Consequently LOFAR will be able to readily
test this unique expectation of the {\it turbulent re-acceleration} model.

\section{Summary \& Conclusions}

In the present paper we perform Monte Carlo simulations to model the formation and evolution of giant radio halos in the framework of the merger-induced particle acceleration scenario (see Sec. 2). Following Cassano et al. (2006a) 
we use homogeneous models that assume {\it a)} an average value of the magnetic field strength in the radio halo volume that scales with cluster mass as $B = B_{<M>} M_v^b$,
and {\it b)} that a fraction, $\eta_t$, of the $P dV$ work, done by subclusters crossing the main clusters during mergers goes into {\it magneto-acoustic} turbulence.
Although simple, these models reproduce the presently observed fraction of galaxy clusters with radio halos and the scalings between the monochromatic radio power of halos at 1.4 GHz and the mass and X-ray luminosity of the host clusters (\eg Cassano et al.~2006, 2008; Venturi et al. 2008), provided that the model parameters $(B_{<M>}, b , \eta_t)$ lie within a fairly constrained range of values (Fig.~7 in Cassano et al. 2006a); in the present paper we adopt a reference set of parameters: $<B>=1.9\, \mu$G, $b=1.5$, $\eta_t=0.2$, that falls in that range.

In Fig.~\ref{fig:NC_NVSS_GMRT} we show that the expected number counts of giant radio halos at $\nu_o=1.4$ GHz obtained with this set of parameters are nicely in agreement with both the data at low redshift (NVSS-XBACS selected radio halos, Giovannini et al.~1999) and at intermediate redshift (clusters in the ``GMRT radio halo survey'', Venturi et al.~2007, 2008).

The most important expectation of the {\it turbulent re-acceleration} scenario
is that the synchrotron spectrum of radio halos should become gradually steeper above a frequency, $\nu_s$ that is determined by the energetics of the merger events that generate the halos and by the electron radiative losses (\eg Fujita et al.~2003; Cassano \& Brunetti 2005).
Consequently, the population of radio halos is expected to be constituted by a mixture of halos with different spectra, with steep spectrum halos being more common
in the Universe than those with flatter spectra (\eg Cassano et al.~2006).
The discovery of these very steep spectrum halos will allow to test the above theoretical conjectures.

In Sect.~2 we have 
derived the expected radio halo luminosity functions
(RHLF) at frequency $\nu_o$, that account for the contribution 
from the different populations of radio halos with $\nu_s \geq \nu_o$. 
The RHLF are obtained combining the theoretical mass function 
of radio halos (with different $\nu_s \geq \nu_o$) with the 
radio power--cluster mass
correlation (Eq.\ref{RHLF}). 
The expected monochromatic radio power at $\nu_o$ of halos hosted 
by clusters with mass $M_v$ is extrapolated from the observed 
$P(1.4)$--$M_v$ correlation by assuming simple scaling relations, appropriate 
for homogenous models, that account for the dependence of the emitted synchrotron power on 
$\nu_s$ (Eqs.\ref{scale2},\ref{scale3}).

\noindent As a relevant case we calculate the expected RHLF at $\nu_o$= 120 MHz
(Fig.~\ref{Fig.RHLF}).
The shape of the RHLF can be approximated by a power law over more than 
two orders of magnitude in radio power. 
Homogeneous models imply the following scalings between $\nu_s$,
cluster mass and the radio luminosity at $\nu_o$, 
$P_{\nu_s} (\nu_o)$ :

\begin{equation}
\nu_s \propto M^{4/3 +b} {{(1 + \Delta M / M)^3}\over
{( <B>^2 + B_{cmb}^2 )^2}}
\label{nus_concl}
\end{equation}

\noindent
and from Eq.~\ref{scale3} and the $P_{1.4}-M_v$ correlation:

\begin{equation}
P_{\nu_s} (\nu_o) \propto M_v^3 \nu_s^{\alpha}
\label{p_concl}
\end{equation}

\noindent
i.e., radio halos with larger $\nu_s$ are typically generated 
in massive clusters that undergo major mergers and contribute
to the RHLF at larger powers. On the other hand halos with smaller
$\nu_s$ are typically generated in less massive systems and
contribute to the RHLF at fainter powers.
Radio halos with $\nu_s \geq$ 120 MHz however become increasingly rare
in clusters with mass $\leq 5 \times 10^{14}$M$_{\odot}$ and this
explains the drop of the RHLF at lower radio powers in Fig.~\ref{Fig.RHLF}.
At the same time, halos with monochromatic radio emission at 
120 MHz $> 10^{26}$W Hz$^{-1}$ would be generated in connection with 
very energetics merging events in very massive clusters, 
that are extremely rare, and this explains the RHLF cut-off at higher 
synchrotron powers in Fig.~\ref{Fig.RHLF}.

In Sect.~3 we discuss the expected number counts of radio halos
at 120 MHz, that best allow us to explore the potential of incoming
LOFAR surveys in constraining present models.

\noindent
The crucial step in this analysis is the estimate of the minimum
diffuse flux from giant radio halos that is detectable by these surveys. 
Because the LOFAR capabilities will become more clear during the incoming 
commissioning phase we exploit two complementary approaches: {\it i)} we required that at 
least half of the radio halo emission is above a fixed brightness--threshold, $\xi\,F$
($F$ being the $rms$ of LOFAR surveys; {\it ii)} we required that 
the signal from the radio halo is $ \geq 3 \times \xi\,F$ in at least 5 beam area of 
LOFAR observations. 
In both cases we assume a fixed shape of the radial profile of radio halos, 
calibrated through that of several well studied halos at 1.4 GHz, 
that introduces a potential source of uncertainty.

Although the uncertainties due to the unavoidable simplifications
in our calculations, the expected number counts of radio halos
highlights the potential of future LOFAR surveys.
By assuming the expected sensitivity of the LOFAR all sky
survey (\eg R\"ottgering 2009; priv. comm.), rms =0.1 mJy/b, and $\xi \sim 2-3$, 
we predict that about 350 giant radio halos ($\sim 200$ considering the case {\it ii)})
can be detected at redshift $\leq$0.6. This means that LOFAR will increase the statistics of these sources by a factor of $\sim 20$ with respect to present day surveys.
About 55\% of
these halos are predicted with a synchrotron spectral
index $\alpha > 1.9$ between 250-600 MHz, and would brighten 
only at lower frequencies, unaccessible to present observations.
Most important, the spectral properties of the population of radio halos
are expected to change with increasing the sensitivity
of the surveys as steep spectrum radio halos
are expected to populate the low-power end of the RHLF. 
A large fraction of radio halos with spectrum
steeper than $\alpha \approx 1.5$ (\eg Fig.~\ref{Fig.alpha}) is expected to allow
a prompt discrimination between different models for the origin of 
radio halos, for instance in this case simple 
energetic arguments would rule out a 
secondary origin of the emitting electrons (\eg Brunetti 2004; Brunetti et al. 2008).

Due to the large number of expected radio halos, 
a potential problem with these surveys is the identification of
halos and of their hosting clusters.
As a matter of fact we expect that LOFAR surveys will open the
possibility to unveil radio halos in galaxy clusters with masses 
$\gtsim 6-7 \times 10^{14}$M$_{\odot}$ at intermediate redshift.
On the other hand, statistical samples of X-ray selected clusters, that
are unique tools for the identification of the hosting clusters,
typically select more massive clusters at intermediate z. 
Consequently, we explored the potential of the first LOFAR surveys as deep
follow ups of available X-ray selected samples of galaxy clusters.

\noindent
We calculate the radio halo number counts expected from 
the follow up of clusters in the eBCS and MACS samples that collect
$\sim 400$ galaxy clusters in the redshift range 0--0.6.
We expect that the LOFAR all sky survey, with a planned sensitivity in
line with the case $\xi\,F=0.25$ mJy/b, will discover about 130 
radio halos in eBCS and MACS clusters and that about 40\% of these radio 
halos should have very steep spectrum, $\nu_s \leq 600$ MHz.
The majority of radio halos in eBCS and MACS clusters are expected
at z = 0.2-0.4, while the small number of clusters at $z \geq 0.5$ in
the MACS catalog does not allow a statistically solid expectations, although we may expect
a couple of radio halos hosted in MACS clusters at this redshift.

The MS$^3$ survey will be carried out in 2010, covering the northern hemisphere, and is expected to reach a noise level of about 0.5 mJy/b at 150 MHz, that implies a sensitivity to diffuse emission from galaxy clusters about one order of magnitude (assuming $\alpha \approx 1.3$)
better than present surveys (\eg NVSS, Condon et al.~1998; 
VLSS, Cohen et al.~2007; WENSS, Rengelink et al.~1997).

\noindent
We considered MS$^3$ pointings relative to the fields of the about 300 galaxy clusters at $z\leq0.3$ in the eBCS catalogues.
We find that about 60 radio halos are expected to be detected by
MS$^3$ observations in these clusters,  
25\% of them (10-15 halos) are expected with $\nu_s \leq$ 600 MHz.
Fairly sensitive GMRT observations of eBCS clusters at 
redshift 0.2--0.3 are already available (Venturi et al.~2007, 2008) and 
we expect that in a few cases radio halos would glow up in the MS$^3$ images
where no diffuse radio emission is detected at 610 MHz. We also find that MS$^3$ observations of eBCS clusters at $z=0-0.2$ can be used to test the increase of the fraction of cluster with radio halos with the X-ray luminosity of the host clusters, which is a unique prediction of our model (Fig.\ref{Fig.histo_Lx}.

The most important simplification in our calculations is the use of homogeneous models. Non-homogeneous approaches, that model the spatial dependence of the acceleration efficiency and magnetic field in the halo volume (\eg Brunetti et al. 2004), and possibly their combination with future numerical simulations, will provide a further step to interpret LOFAR data. Also the use of the extended PS theory is expected to introduce some biases. For instance, it is well-known that the PS mass function underproduces the expected number of massive clusters ($M>10^{15}\,M_{\odot}$) at higher redshift, $z\sim 0.4-0.5$, by a factor of $\sim 2$ with respect to that found in N-body simulations (\eg Governato et al. 1999; Bode et al. 2001; Jenkins et al. 2001).
Since in our model the great majority of halos at these redshift is associated with massive clusters, the use of the PS mass function implies that the RHNC at $z> 0.4-0.5$ could be underestimated by a similar factor. A further refinement of the approach proposed in the present paper could be obtained with the use of galaxy cluster {\it merger trees} extracted from N-body simulations. These would also allow for a more {\it realistic} description of the merger events (spatially resolved, multiple mergers, etc...).
 
In the present paper we focus on a reference set of model parameters. Cassano et al. (2006a) discussed the dependence of model expectations at 1.4 GHz on these parameters. Based on their analysis we expect that all the general results given in the present paper are independent of the adopted values for parameters. The expected number counts of halos should change by a factor of $\sim2-2.5$ considering sets of model parameters within the region ($<B>$, $b$, $\eta_t$) that allows for reproducing the observed $P_{1.4}-M_v$ correlation. In this case the number of halos that we expect decreases from super-linear sets of parameters ($b>1$ and $<B>\geq 1.5\,\mu$G) to sub-linear cases ($b<1$ and $<B>\leq 1.5\,\mu$G) (see also Fig.4 in Cassano et al. 2006b); a more detailed study will be presented in a future paper.

\begin{acknowledgements}
We thank the anonymous referee for useful comments. This work is partially supported by ASI and INAF under grants PRIN-INAF 2007, PRIN-INAF 2008 and ASI-INAF I/088/06/0. 
\end{acknowledgements}


\begin{thebibliography}{}

\bibitem{} Bacchi M., Feretti L., Giovannini G., Govoni F., 2003, A\&A, 400, 465
\bibitem{} Blasi P., 2001, APh 15, 223
\bibitem{} Blasi, P., \& Colafrancesco, S.\ 1999,
\bibitem{} Bode P, Bahcall N.A., Ford E. B. \& Ostriker J.P., 2001, ApJ 551, 15
\bibitem{} Bonafede A., Feretti L., Giovannini G., Govoni F., Murgia M., Taylor G. B., Ebeling H, Allen S., Gentile G., Pihlstrom Y., 2009, A\&A in press; arXiv:0905.3552
\bibitem{} B\"ohringer, H.; Voges, W.; Huchra, J. P.; et al. 2000, ApJS 129, 435
\bibitem{} B{\"o}hringer, H., Schuecker, P., Guzzo, L., et al., 2004, A\&A, 425, 367
\bibitem{} Brunetti G., 2004, JKAS 37, 493
\bibitem{} Brunetti G., Setti G., Feretti L., Giovannini G., 2001, MNRAS 320, 365
\bibitem{} Brunetti G., Blasi P., Cassano R., Gabici S., 2004, MNRAS 350, 1174
\bibitem{} Brunetti G., Lazarian A., 2007, MNRAS 378, 245
\bibitem{} Brunetti, G.; Venturi, T.; Dallacasa, D., Cassano, R., Dolag, K., Giacintucci, S., Setti, G., 2007, ApJ, 670, 5
\bibitem{} Brunetti, G., Giacintucci, S., Cassano, R., Lane, W., Dallacasa, D., Venturi, T., Kassim, N. E., Setti, G., Cotton, W. D., Markevitch, M., 2008, {\it Nature}, 455, 944
\bibitem{} Brunetti, Cassano, Dolag, Setti 2009, A\&A in press, arXiv:0909.2343 
\bibitem{} Br\"uggen, M., Ruszkowski, M., Simionescu, A., Hoeft, M., Dalla Vecchia, C., 2005, ApJ, 631, L21
\bibitem{} Buote D.A, 2001, ApJ 553, 15
\bibitem{} Cassano R. \& Brunetti G., 2005, MNRAS 357, 1313 
\bibitem{} Cassano R., Brunetti G., Setti G., 2006a, MNRAS 369,1577 
\bibitem{} Cassano R., Brunetti G., Setti G., 2006b, Astronomische Nachrichten 327, 557
\bibitem{} Cassano, R.; Brunetti, G.; Venturi, T., Setti, G., Dallacasa, D., Giacintucci, S., Bardelli, S., 2008, A\&A, 480, 687
\bibitem{} Cohen, A. S.; Lane, W. M.; Cotton, W. D.; Kassim, N. E.; Lazio,
T. J. W.; Perley, R. A.; Condon, J. J.; Erickson, W. C., 2007, AJ 134, 1245
\bibitem{} Condon J.J., Cotton W.D., Greisen E.W., Yin Q.F., Perley R.A., Taylor G.B., Broderick J.J., 1998, AJ 115, 1693
\bibitem{} Dallacasa, D.; Brunetti, G.; Giacintucci, S.; Cassano, R.; Venturi, T.; Macario, G.; Kassim, N. E.; Lane, W.; Setti, G., 2009, ApJ 699, 1288	
\bibitem{} Dennison B., 1980, ApJ 239
\bibitem{} Dolag K., Bartelmann M., Lesch H, 2002, A\&A, 387, 383
\bibitem{} Ebeling H., Voges W., Bohringer H., Edge A.~C., Huchra J.~P., \& Briel U.~G.\ 1996, MNRAS 281, 799 
\bibitem{} Ebeling H., Edge A.C., B\"ohringer H., et al.,1998 MNRAS, 301, 881
\bibitem{} Ebeling H., Edge A.C., Allen S.W., et al., 2000, MNRAS 318, 333
\bibitem{} Ebeling, H.; Edge, A. C.; Henry, J. P., 2001, ApJ, 553, 668
\bibitem{} Ellingson S.W., Clarke T.E., Cohen A., Craig J., Kassim N.E., Pihlstrom Y., Rickard L.J., Taylor G.B., Proc. of the IEEE, Vol. 97, Issue 8, pp 1421-1430
\bibitem{} En\ss lin T.~A., Biermann P.~L., Klein U., \& Kohle S.\ 1998, \aap, 332, 395 
\bibitem{} En\ss lin T.A., R\"ottgering H., 2002, A\&A, 396, 83
\bibitem{} Feretti L., 2005, in 'X-Ray and Radio Connections', published electronically by NRAO, eds. L.O.Sjouwerman and K.K.Dyer
\bibitem{} Ferrari, C., Govoni, F., Schindler, S., Bykov, A. M., Rephaeli, Y., 2008, Space Science Reviews, 134, 93
\bibitem{} Fujita Y., Takizawa M., Sarazin C.L., 2003, ApJ 584, 190
\bibitem{} Giovannini G., Tordi M., Feretti L., 1999, NewA 4, 141, (GTF99)
\bibitem{} Governato, F.; Babul, A.; Quinn, T.; Tozzi, P.; Baugh, C. M.; Katz, N.; Lake, G., 1999, MNRAS 307, 949
\bibitem{} Govoni F., Markevitch M., Vikhlinin A., VanSpeybroeck L., Feretti, L., Giovannini G., 2004, ApJ 605, 695
\bibitem{} Govoni F., Feretti L., Giovannini G., B\"oringer H., Reiprich T.H., Murgia M., 2001, A\&A, 376, 803
\bibitem{} Hwang C.-Y., 2004, JKAS 37, 461
\bibitem{} Hoeft, M., Br\"uggen, M., 2007, MNRAS, 375, 77
\bibitem{} Hoeft, M.; Br\"uggen, M.; Yepes, G.; Gottlöber, S.; Schwope, A., 2008, MNRAS 391, 1511
\bibitem{} Jenkins, A.; Frenk, C. S.; White, S. D. M.; Colberg, J. M.; Cole, S.; Evrard, A. E.; Couchman, H. M. P.; Yoshida, N., 2001, MNRAS, 321, 372
\bibitem{} Liang H., Hunstead R.W., Birkinshaw M., Andreani P., 2000, ApJ 544, 686
\bibitem{} Lacey, C., \& Cole, S.\ 1993, \mnras, 262, 627 
\bibitem{} Miniati F., Jones T.W., Kang H., Ryu D., 2001, \apj 562, 233
\bibitem{} Petrosian V., 2001, ApJ 557, 560 
\bibitem{} Pfrommer, C., En\ss lin, T. A., Springel, V., 2008, MNRAS, 385, 1211
\bibitem{} Press W.H., Schechter P., 1974, ApJ 187, 425
\bibitem{} Rengelink, R. B.; Tang, Y.; de Bruyn, A. G.; Miley, G. K.;
Bremer, M. N.; R\"ottgering, H. J. A.; Bremer, M. A. R., 1997, A\&AS 124, 259
\bibitem{} R\"ottgering, H. J. A.; Braun, R.; Barthel, P. D., et al. 2006, proceedings of the conference ``Cosmology, galaxy formation and astroparticle physics on the pathway to the SKA", Oxford, April 10-12 2006, astro-ph/0610596
\bibitem{} Ryu, D., Kang, H., Hallman, E., Jones, T.W., 2003, ApJ 593, 599
\bibitem{} Ryu, D., Kang, H., Cho, J., Das, S., 2008, Science, 320, 909
\bibitem{} Sarazin C.L., 1999, ApJ 520, 529
\bibitem{} Schlickeiser R., Sievers A., Thiemann H., 1987, A\&A 182, 21
\bibitem{} Schlickeiser, R.; Miller, J. A., 1998, ApJ 492, 352
\bibitem{} Schuecker P., B\"ohringer H.; Reiprich T.H., Feretti L., 2001, A\&A 378, 408
\bibitem{} Springel, V. et al. 2005, {\it Nature}, 435, 629
\bibitem{} Subramanian, K., Shukurov, A., Haugen, N. E. L., 2006, MNRAS 366, 1437
\bibitem{} Thierbach, M., Klein, U., \& Wielebinski, R.\ 2003, \aap, 397, 53
\bibitem{} Tr\"umper, J., 1993, {\it Science}, 260, 1769
\bibitem{} Venturi T., Giacintucci S., Brunetti G., Cassano R., Bardelli S., Dallacasa D., Setti G., 2007, A\&A 463, 937
\bibitem{} Venturi, T., Giacintucci, S., Dallacasa, D., Cassano, R., Brunetti, G., Bardelli, S., Setti, G., 2008, A\&A, 484, 327
\bibitem{} van Weeren R.J., R\"ottgering H.J.A., Bruggen M, Cohen A., 2009, A\&A in press;  arXiv:0905.3650
\end{thebibliography}
\end{document}